\documentclass[%
 reprint,
 amsmath,amssymb,
 aps,
]{revtex4-2}

\usepackage{graphicx}
\usepackage{dcolumn}
\usepackage{bm}
\usepackage{color}

\begin{document}

\preprint{APS/123-QED}

\title{Properties of magnetohydrodynamic modes in compressively driven plasma turbulence}

\author{K. D. Makwana$^{1}$}
 \email{kirit.makwana@desy.de}
\author{Huirong Yan$^{1,2}$}%
 \email{huirong.yan@desy.de}
\affiliation{%
 $^{1}$Deutsches Elektronen Synchrotron (DESY), Platanenallee 6, D-15738 Zeuthen, Germany\\
 $^{2}$Institut f{\"u}r Physik und Astronomie, Universit{\"a}t Potsdam, D-14476 Potsdam, Germany
}%


%

\date{\today}

\begin{abstract}
We study properties of magnetohydrodynamic (MHD) eigenmodes by decomposing the data of MHD simulations into linear MHD modes - namely the Alfv{\'e}n, slow magnetosonic, and fast magnetosonic  modes. We drive turbulence with a mixture of solenoidal and compressive driving, while varying the Alfv{\'e}n Mach number ($M_A$), plasma $\beta$, and the sonic Mach number from sub-sonic to trans-sonic. We find that the proportion of fast and slow modes in the mode mixture increases with increasing compressive forcing. This proportion of the magnetosonic modes can also  become the dominant fraction in the mode mixture. The anisotropy of the modes is analyzed by means of their structure functions. The Alfv{\'e}n mode anisotropy is consistent with the Goldreich-Sridhar theory. We find a transition from weak to strong Alfv{\'e}nic turbulence as we go from low to high $M_A$. The slow mode properties are similar to the Alfv{\'e}n mode. On the other hand the isotropic nature of fast modes is verified in the cases where the fast mode is a significant fraction of the mode mixture. The fast mode behavior does not show any transition in going from low to high $M_A$. We find indications that there is some interaction between the different modes and the properties of the dominant mode can affect the properties of the weaker modes. This work  identifies the conditions under which magnetosonic modes can be a major fraction of turbulent astrophysical plasmas, including the regime of weak turbulence. Important astrophysical implications for cosmic ray transport and magnetic reconnection are discussed.
\end{abstract}

\maketitle


\section{\label{introduction}Introduction}

Plasma turbulence plays an important role in various astrophysical processes. It is important in solar wind heating and acceleration~\cite{BrunoCarbone2013}, it regulates star formation processes~\cite{MacLowKlessen2004,PadoanFederrath2014,Federrath2018}, and it scatters cosmic rays~\cite{Schlickeiser2002} amongst other things. The properties of turbulence depend on the underlying modes it is made up of. The magnetohydrodynamic (MHD) system of equations of a 3D, homogenous, uniform, isothermal plasma with a uniform background magnetic field allows for three separate propagating linear eigenmodes - the Alfv{\'e}n mode~\cite{Alfven1942}, the slow magnetosonic mode, and the fast magnetosonic mode~\cite{Swanson2003}. Alfv{\'e}nic turbulence (turbulence consisting of mostly Alfv{\'e}n modes interacting with each other) is thought to be quite important in solar turbulence as Alfv{\'e}n waves have been observed in the solar wind~\cite{BelcherDavis1971}. Alfv{\'e}nic turbulence has been studied for several decades and several theories have been developed to describe it. The Alfv{\'e}n modes are incompressible solutions to the linearized MHD equations. In the regime of strong turbulence, a critical balance is conjectured to be reached between the linear interaction time of wave packets with their nonlinear cascade time. As a result, scale dependent anisotropy appears \cite{GoldreichSridhar1995}. In the limit of weak turbulence, the resonant three-wave couplings involve only the non-propagating Alfv{\'e}n modes and produce a cascade in the wavevectors perpendicular to the local mean magnetic field direction only~\cite{ShebalinMatthaeus1983}. 

There has not been a comprehensive theory of turbulence consisting of the compressible MHD modes, namely the slow and fast magnetosonic modes. Turbulence in the interstellar medium is identified by the measurement of density fluctuations in it, indicating the presence of compressible turbulence~\cite{ArmstrongRickett1995}. These turbulent density fluctuations in the interstellar medium molecular clouds are closely linked to star formation~\cite{Larson1981,RomanDuvalFederrath2011,FederrathRathborne2016}. Numerical simulations of compressible turbulence to identify the inertial range scalings are more difficult and complex, with varying results depending on the plasma parameters like Mach number~\cite{VestutoOstriker2003, Federrath2013, GreteOshea2017}. Since astrophysical turbulence is expected to be magnetized and compressible, the magnetosonic modes should be considered when studying such turbulence. Whether the cascades of these different MHD modes are independent of each other and what is their nature are still open questions. The slow modes are cascaded by the shear-Alfv{\'e}n modes and hence are expected to behave like the Alfv{\'e}n modes~\cite{LithwickGoldreich2001}. Earlier studies indicated that the interactions between the Alfv{\'e}n and the magnetosonic modes are limited on scales smaller than the injection scale. They also have shown that the energy spectrum and anisotropy of slow modes is quite similar to Alfv{\'e}n modes~\cite{ChoLazarian2003}. On the other hand, fast modes have been seen to be quite different in their spectrum and anisotropy characteristics. Unlike Alfv{\'e}n modes which preferentially cascade in the field perpendicular direction, fast modes seem to show an isotropic cascade. This has also led to several important implications for astrophysical turbulence. Based on this it has been shown that fast modes could be the most effective scatterers of cosmic ray particles~\citep{YanLazarian2002, YanLazarian2004}. 

Particle scattering and diffusion critically depends on the properties of plasma turbulence. While fast modes can play an important role in scattering of cosmic rays, simulations have shown that the fast modes might only be a marginal component of compressible turbulence~\cite{YangZhang2018}. However, these simulations have been driven incompressively by solenoidal forcing~\citep{VestutoOstriker2003, ChoLazarian2003, YangZhang2018, KowalLazarian2007,AndresLeoni2017}. So a natural question to ask is whether and how the nature of the forcing affects the mode composition of turbulence. We try to answer this by compressively driving turbulence in a variety of different plasma parameter regimes. Some earlier studies have driven turbulence by keeping a mixture of solenoidal and compressive velocity field at large scales~\citep{YangShi2016} or by decomposing the driving force into solenoidal and compressive components~\citep{FederrathRomanDuval2010}. We adopt a similar forcing but focus on the MHD mode decomposition. We find that the nature of the forcing significantly affects the composition of the turbulence in terms of the MHD modes. Another phenomenon less explored in numerical simulations is low $M_A$ (ratio of r.m.s velocity to Alfv{\'e}n speed) turbulence. Theoretically, the Alfv{\'e}nic turbulent cascade is expected to be weak up to some scale and then transition to a state of strong turbulence, mediated by the critical balance condition, at smaller scales. This transition from weak to strong turbulence has only been recently simulated in decaying turbulence~\citep{MeyrandGaltier2016}. We explore the $M_A$ dependence of this transition in our compressively driven simulations and find that the nature of Alfv{\'e}nic turbulence can significantly change with $M_A$.

The isotropic cascade properties of the fast mode have been studied at limited resolution previously~\citep{ChoLazarian2003} and the spectrum of this cascade was cautiously claimed to be $k^{-3/2}$. Higher resolution studies are needed to verify this behavior through a well-resolved inertial range. We perform higher resolution studies and find the isotropic nature of the fast modes throughout the inertial range, suggesting no scale-dependent anisotropy. Another related question is whether fast modes also show an $M_A$ dependent behavior like Alfv{\'e}n modes in terms of weak or strong turbulence? We do not find any such dependence for the fast modes. Our results also suggest that these different modes cascade are not completely independent of each other, depending on which mode is dominant. This study shows that the nature of MHD turbulence can be different depending on a variety of parameters, particularly, the nature of driving, and this has important implications for understanding the effect of this turbulence on related problems. 

\section{\label{sec2}Simulation setup and mode decomposition}

The simulations are performed by using the PLUTO code~\cite{MignoneBodo2007}. The ideal MHD equations are solved with no explicit resistivity or viscosity, only with numerical dissipation. The isothermal equation of state is used. The in-built HLLD Riemann solver~\cite{Mignone2007} is utilized in conjunction with a WENO3 reconstruction scheme~\cite{YamaleevCarpenter2009}. The time stepping is done by a $3^{\text{rd}}$ order Runge-Kutta scheme. The simulation box is a cube of length $L_x=L_y=L_z=1$. The normalization is such that the Alfv{\'e}n velocity $v_A$ and the mean magnetic field $B_0$ are numerically same. The mean field $\bm{B}_0$ is in the $z$ direction which is the global parallel direction. The dimensionless quantities like plasma $\beta$ and Mach numbers are given in Table.~\ref{tab:table1}. The sound speed ($c_s$) is changed to vary the plasma $\beta$ defined as $\beta\equiv2(c_{s}/v_{A})^2$. 

Turbulence is driven by using a readily available forcing module in PLUTO. This drives turbulence by adding a force $\bm{F}^{turb}$  in both the momentum and energy equations. This force is modeled as an Ornstein-Uhlenbeck (OU) process~\cite{EswaranPope1988,FederrathKlessen2008}. This process is a stochastic differential equation governing the evolution of the force $\bm{F}^{turb}$, given by,
\begin{equation}
d\bm{F}^{\text{turb}}(\bm{k},t)=\bm{F}_{0}^{\text{turb}}(\bm{k})\mathcal{P}^{\zeta}(\bm{k})d\mathcal{W}(t)-\bm{F}^{\text{turb}}(\bm{k},t)\frac{dt}{T}
\end{equation}
Here $d\bm{F}$ is a force added at every time step to the existing force $\bm{F}^{\text{turb}}(\bm{k})$ at wavevector $\bm{k}$. The $\mathcal{P}^{\zeta}(\bm{k})$ operator is a projection operator which separates the solenoidal and compressive parts of the force.
\begin{equation}
\mathcal{P}^{\zeta}_{ij}(\bm{k})=\zeta\mathcal{P}_{ij}^{\perp}(\bm{k})+(1-\zeta)\mathcal{P}_{ij}^{\parallel}(\bm{k})=\zeta\delta_{ij}+(1-2\zeta)\frac{k_ik_j}{k^2}
\end{equation}
When $\zeta=1$ the forcing is purely solenoidal, when $\zeta=0$ it is purely compressive, and intermediate values give a mixture of solenoidal and compressive forcing. The forcing is limited to a range of wavenumbers $k_{\text{min}} \le k \le k_{\text{max}}$. We ignore the $2\pi$ factor in the definition of the wavenumber and since the box length is also normalized to unity, the smallest wavenumber possible in our simulation is 1. Thus $k_{\text{min}}=1$ and $k_{\text{max}}=3$ in our simulations. We vary the Mach number $M_A$ of the turbulence by varying the energy injected into the turbulence $E_{\text{inj}}$. We also vary the plasma $\beta$ by changing the isothermal sound speed $c_s$. This changes the sonic Mach number also as shown in Table.~1 for the different simulations. We scan the sonic Mach number from sub-sonic to trans-sonic range. We utilize solenoidal driving ($\zeta=1$) and mostly compressive driving ($\zeta=0.1$). Two different grid resolutions are utilized, $512^3$ and $1024^3$. Table.~1 shows the different simulations we have analyzed and their parameters. The first letter ``S" or ``C" in the simulation ID represents whether it is solenoidally or compressively driven. The letters ``a" or ``b" denotes the resolution ($512^3$ or $1024^3$ respectively). Table.~1 also lists the correlation time of the forcing $T$. This is close to the usual value of eddy turnover time at the injection scale ($L_{inj}$), $T\approx (L_{inj})/v \approx (1/2)/(v_AM_A)$~\cite{SchmidtHillebrandt2006}. The simulations are run for several 10's of eddy correlation time $T$ so that there are many snapshots of a statistical steady state. 

\begin{table}[h]
\caption{\label{tab:table1}%
Simulation parameters in steady state with simulation IDs. The energy injection rate $E_{inj}$, the plasma $\beta$, Alfv{\'e}n Mach number $M_A$, sonic Mach number $M_S$, the forcing correlation time $T$, resolution, and fraction of compressive driving $\zeta$ is varied amongst the different simulation runs.}
\begin{ruledtabular}
\begin{tabular}{cccccccc}
{ID} & {$E_{\text{inj}}$} & {$\beta$} & {$M_A$} & {$M_S$} & T & {Resolution} & {$\zeta$}\\
\colrule
S1a & $10^{-8}$ & 2.17 & 0.24 & {0.23} & 20 &$512^3$ & 1.0 \\
S2a & $8\times 10^{-8}$ & 2.17 & 0.46 & {0.44} & 10 & $512^3$ & 1.0 \\
{S2.5a} & {$2\times 10^{-7}$} & {2.17} & {0.59} & {0.56} & {8.5} & {$512^3$} & 1.0 \\
S3a & $5\times 10^{-7}$ & 2.17 & 0.69 & {0.66} & 7.5 & $512^3$ & 1.0 \\
S4a & $8\times 10^{-6}$ & 2.17 & 0.99 & {0.95} & 5 & $512^3$ & 1.0 \\
C1a & $3\times 10^{-7}$ & 2.17 & 0.22 & {0.21} & 20 & $512^3$ & 0.1 \\
C2a & $5\times 10^{-6}$ & 2.17 & 0.48 & {0.46} & 10 & $512^3$ & 0.1 \\
{C3a} & {$2 \times 10^{-5}$} & {2.17} & {0.66} & {0.63} & {7.5} & {$512^3$} & {0.1} \\
C4a & $9\times 10^{-5}$ & 2.17 & 1.03 & {0.99} & 5 & $512^3$ & 0.1 \\
CB0a & $8\times 10^{-6}$ & 0.5 & 0.51 & {1.02} & 10 & $512^3$ & 0.1 \\
CB1a & $3\times 10^{-6}$ & 8.0 & 0.60 & {0.3} & 10 & $512^3$ & 0.1 \\
S1b & $10^{-8}$ & 2.17 & 0.25 & {0.24} & 20 &$1024^3$ & 1.0 \\
S2b & $8\times 10^{-8}$ & 2.17 & 0.48 & {0.46} & 10 & $1024^3$ & 1.0 \\
S3b & $5\times 10^{-7}$ & 2.17 & 0.72 & {0.69} & 7.5 & $1024^3$ & 1.0 \\
S4b & $8\times 10^{-6}$ & 2.17 & 0.87 & {0.84} & 5 & $1024^3$ & 1.0 \\
C1b & $3\times 10^{-7}$ & 2.17 & 0.23 & {0.22} & 20 & $1024^3$ & 0.1 \\
C4b & $9\times 10^{-5}$ & 2.17 & 1.05 & {1.01} & 5 & $1024^3$ & 0.1 \\
\end{tabular}
\end{ruledtabular}
\end{table}

The mode decomposition is a linear eigenmode decomposition where the MHD state vector comprising of the density, velocity, and magnetic fluctuations is decomposed into a linear combination of the two Alfv{\'e}n mode eigenvectors, the two fast magnetosonic mode eigenvectors and the two slow magnetosonic mode eigenvectors. The MHD description is generally regarded valid on scales much larger than the typical kinetic scales like {mean free path}, ion skin-depth, and gyro-radius. Although kinetic damping can affect the compressible modes on {collisionless} scales~\cite{ToldCookmeyer2016,KleinHowes2012}, in typical {warm} ISM plasmas there is a vast range of scales from the injection (a few tens -100 pcs) to {collisionless} scales such that MHD prescription is justified for a large range of scales. The linear MHD mode decomposition assumes a homogeneous plasma without strong gradients. It is applicable if the turbulence driving scale is $\leq$ the scale length of the gradients. The mode decomposition is valid in those regions where the background plasma is devoid of strong gradients or discontinuities and where the driving can also be considered homogeneous. In-situ observations of the solar wind have revealed presence of the Alfv{\'e}n and slow modes~\cite{ShiXiao2015}. If we consider the intra-galactic media with a scale height of a few hundred pc, then regions $\lesssim100$pc would be homogeneous in the absence of strong discontinuities. Many observations also indicate that turbulence is driven on galactic-scales with the injection scale $\sim 100$pc~\cite{FalcetaGoncalvesKowal2014}. In such regions with sizes smaller than the injection scales the turbulence driving can be considered homogeneous and the linear MHD mode analysis should apply.

We follow the prescription of Ref.~\citep{ChoLazarian2003} to decompose the MHD data into the MHD eigenmodes. The sonic and Alfv{\'e}n Mach numbers are kept $\lesssim 1$ so that the nonlinear terms ($\delta\bm{v}^2,\delta\bm{b}^2$) are not stronger than the linear terms ($v_A\delta\bm{v},B_0\delta\bm{b}$). In this case the linear mode decomposition will be meaningful. After Fourier transforming the MHD data, the MHD state at each wavevector $\bm{k}$ is arranged in a column vector consisting of $(\rho_{\bm{k}},\bm{v}_{\bm{k}},\bm{b}_{\bm{k}})$, where $\rho$ is the density, $\bm{v}$ is the perturbed velocity field vector (normalized to the Alfv{\'e}n speed), and $\bm{b}$ is the perturbed magnetic field vector (normalized to the mean field $\bm{B}_0$). Here the mean density $\rho_0$ and mean magnetic field $\bm{B}_0$ have been subtracted and only the fluctuating components are kept. The magnetic field components are not truly independent due to the divergence free condition $\bm{k}\cdot\bm{b}_{\bm{k}}=0$. Therefore only 2 components of the $\bm{b}$ field are kept to reduce the MHD state vector to 6 components. These two components are selected to be closest to the magnetic field vector of the Alfv{\'e}n mode and the slow (or fast) mode. 

The 6 eigenmode vectors are placed in a $6\times 6$ matrix $A$, while the MHD state is a column vector $b$, solving for the linear amplitudes, $x$, of the modes by solving the linear matrix equation $Ax=b$. Once the Fourier amplitudes of the different modes are obtained in this manner, an inverse Fourier transform gives us data cubes in the physical space consisting entirely of single, specific MHD modes. The perpendicular propagation, $k_{z}=0$, is a special case where the Alfven and slow modes become degenerate and non-propagating. The parallel propagation case ($k_{\perp}=0$) is also special as here the Alfv{\'e}n mode is degenerate with either the slow or fast mode depending on whether $v_A < c_s$ or $v_A > c_s$. 

Using this decomposition the density, velocity and magnetic fields obtained from MHD simulations are decomposed into three modes, i.e., $\bm{v}_{\bm{k}}=\bm{v}_{\bm{k},A}+\bm{v}_{\bm{k},S}+\bm{v}_{\bm{k},F}$ and $\bm{b}_{\bm{k}}=\bm{b}_{\bm{k},A}+\bm{b}_{\bm{k},S}+\bm{b}_{\bm{k},F}$, where the subscripts $A,S,F$ refer to the Alfv{\'e}n, slow, and fast modes respectively. To measure the relative presence of the different modes in the turbulence, the relative fraction of ``energy" in that mode is calculated by taking, 
\begin{align}
P_{MEm}=100\%\times\frac{\sum_{\bm{k}}|\bm{b}_{\bm{k},m}|^2}{\sum_{m,\bm{k}}(|\bm{b}_{\bm{k},m}|^2+|\bm{v}_{\bm{k},m}|^2)},\label{eqME}\\
P_{KEm}=100\%\times\frac{\sum_{\bm{k}}|\bm{v}_{\bm{k},m}|^2}{\sum_{m,\bm{k}}(|\bm{b}_{\bm{k},m}|^2+|\bm{v}_{\bm{k},m}|^2)}\label{eqKE}.
\end{align}
The variable $m$ represents the MHD modes, standing for $A$, $S$, or $F$. $P_{MEm} ($Eq.~\ref{eqME}) stands for magnetic energy fractions and $P_{KEm}$ (Eq.~\ref{eqKE}) stands for kinetic energy fractions. The sum over $m$ in the denominator means summing over all three modes. The sum over the Fourier modes excludes the $k_{z}=0$ and $k_{\perp}=0$ modes to count only non-degenerate modes. As an example, $P_{MEA}$ stands for the percentage of magnetic energy in Alfv{\'e}n modes while $P_{KEF}$  estimates the fraction of fast mode velocity fluctuations in the total mode mixture. 

\begin{figure}[h]
\centering 
\includegraphics[width=1.0\linewidth]{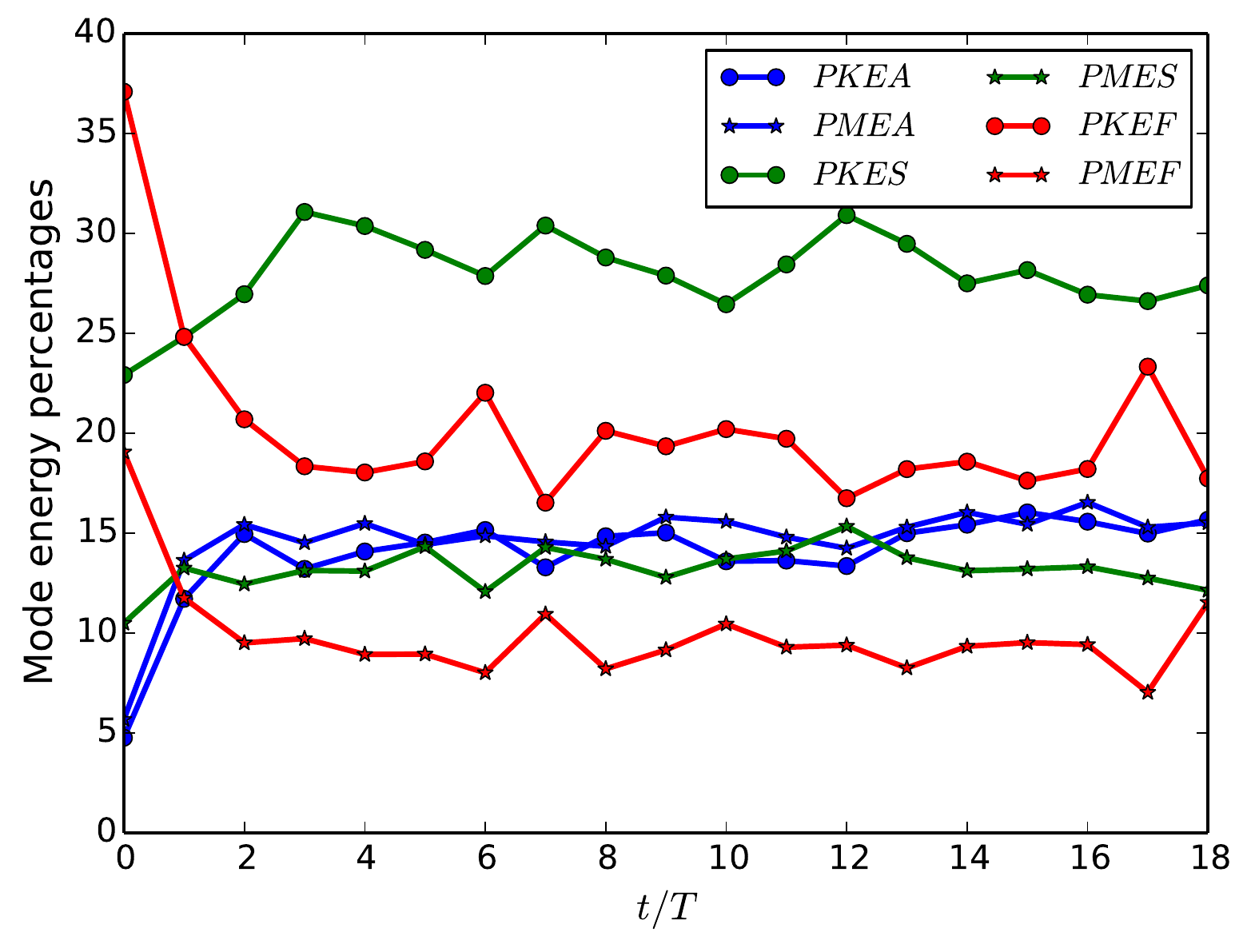}
\caption{Fraction of mode energies in the velocity and magnetic fields of the different MHD modes for the C1a simulation as a function of time.}
\label{e_fracs}
\end{figure}
These mode energy fractions change as a function of time as shown in Fig.~\ref{e_fracs}. It shows the energy fractions as a function of time in the simulation C1a. This simulation is first run at a lower resolution of $128^3$ in order to reach a steady state in energy quickly. Then the $512^3$ simulation is launched using a data cube from the $128^3$ simulation as the initial condition with trilinear interpolation. We see that it takes some time initially for the different mode fractions to attain a steady value. The Alfv{\'e}n mode shows a very similar level of its kinetic and magnetic fluctuations, as expected from its eigenmodes. The dominant contribution comes from the kinetic component of the slow modes, while its magnetic component is comparable to the Alfv{\'e}n modes. The fast mode also shows a strong kinetic component and a weaker magnetic component.

\begin{figure}[h]
\centering 
\includegraphics[width=1.0\linewidth]{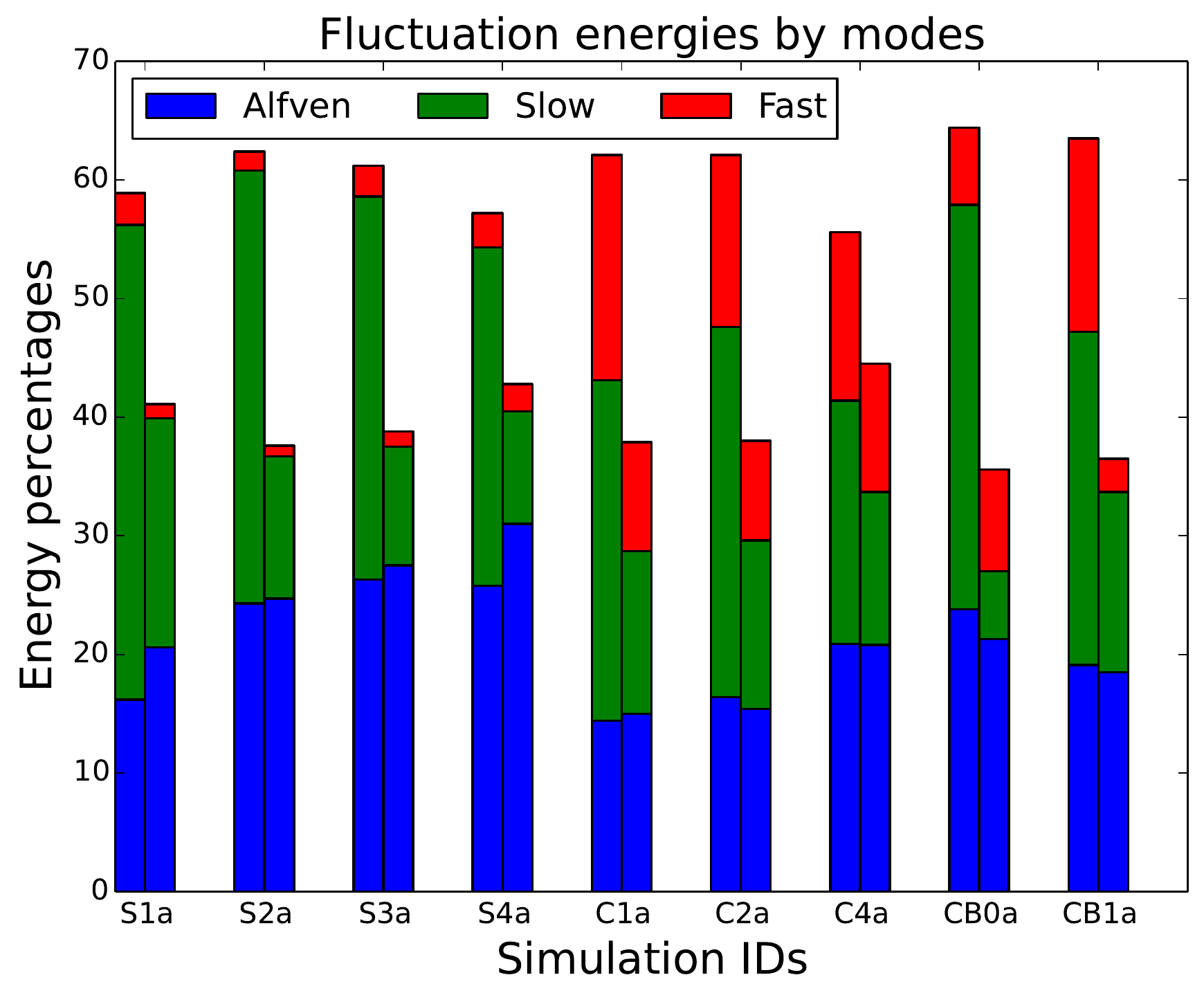}
\caption{The time-averaged fractions of mode energies in different modes in different simulations. Each simulation has two bars, the left one represents the velocity field showing the three mode percentages ($P_{KEA}$, $P_{KES}$, and $P_{KEF}$ in blue, green, and red respectively). Similarly the right hand bar is for the magnetic field showing $P_{MEA}$, $P_{MES}$, and $P_{MEF}$ in their respective colors. Both the bars add up to 100\%. Compressive driving leads to a significantly larger fraction of the fast magnetosonic mode.}
\label{energy_fracs_barplot}
\end{figure}

\begin{figure*}[t]
\centering  
\includegraphics[width=1.0\linewidth]{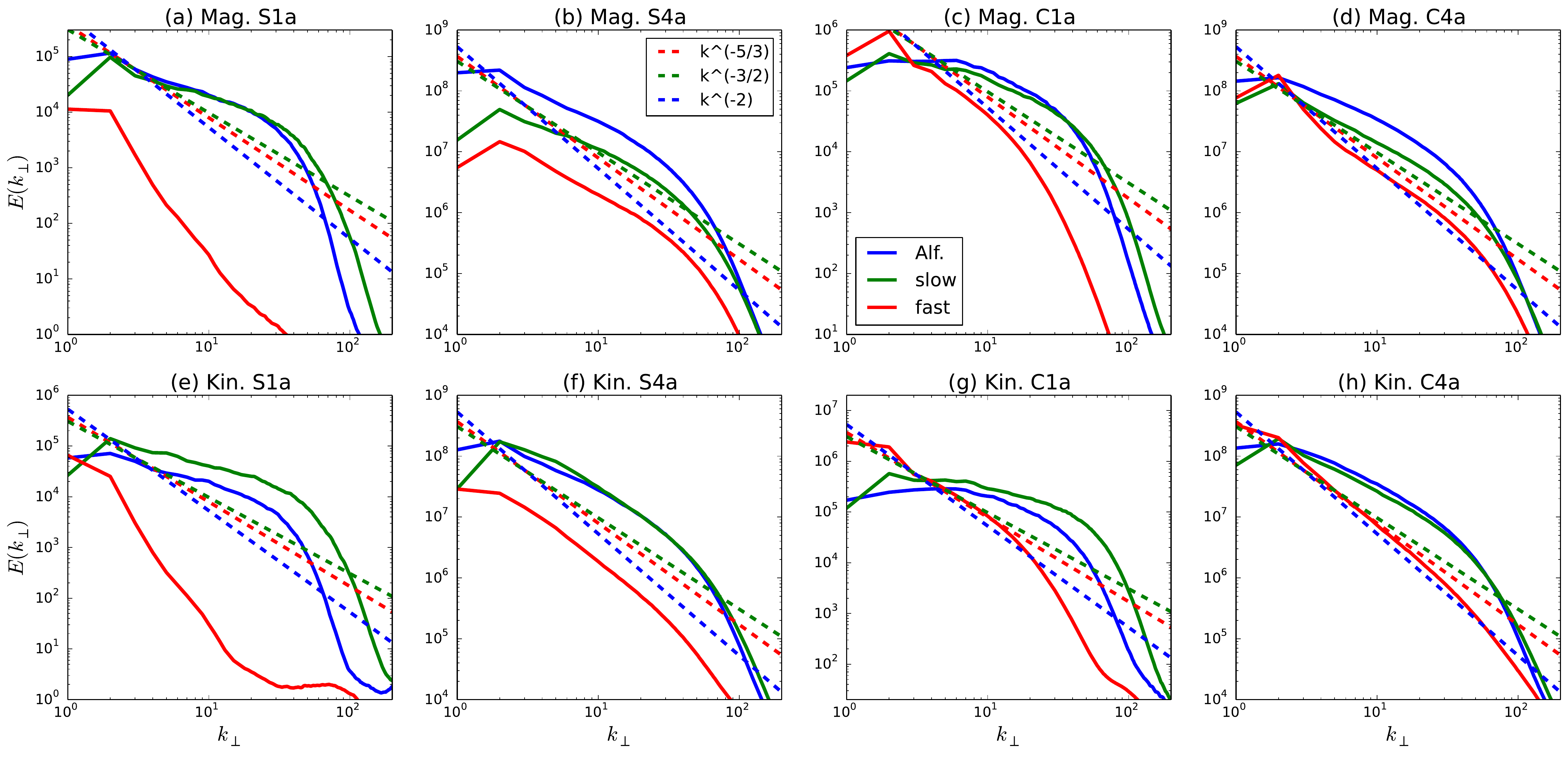}
\caption{Energy spectra of the different MHD modes showing a comparison between S1a, S4a, C1a, and C4a simulations. Top row is spectrum of magnetic fields, bottom row shows the velocity field spectrum. Three reference slopes are also plotted giving $k_{\perp}^{-3/2}$, $k_{\perp}^{-5/3}$, and $k_{\perp}^{-2}$ slopes. The legend applies to all the subplots.}
\label{1Dkspec_array}
\end{figure*}

These mode energy fractions are averaged over the steady state snapshots of each $512^3$ simulation and shown in Fig.~\ref{energy_fracs_barplot}. In the solenoidally driven simulations, the Alfv{\'e}n and slow modes form the major fraction, with very little contribution from fast modes. In simulation S4a the fast mode contributes only 5.1\% to kinetic energy fluctuations and 5.4\% to magnetic energy fluctuations. Going from S1 to S4 as the Alfv{\'e}n Mach number increases, there is a slight increase in the fraction of Alfv{\'e}n modes. The Alfv{\'e}n modes have roughly equal energies in the velocity and magnetic fields while the slow mode has a stronger component of velocity fields. A striking feature of Fig.~\ref{energy_fracs_barplot} is that the fast mode has a significantly large proportion in the compressively driven simulations, which has not been observed before. In simulation C1a the fast mode contributes 30.6\% to kinetic energy fluctuations and 24.3\% to magnetic energy fluctuations. In simulation C4a the fast mode contributes 25.5\% to kinetic energy fluctuations and 24.3\% to magnetic energy fluctuations. On the other hand, comparing C1a, C2a, and C4a shows that changing the $M_A$ is not affecting the mode fractions significantly (except for a gradual increase in Alfv{\'e}n mode proportion). Comparing CB0a with CB1a  shows that kinetic fluctuations of slow modes decrease while their magnetic fraction increases as the plasma $\beta$ increases, taking the slow mode magnetic and kinetic fluctuations closer to equipartition. This is understood from the fact that as $\beta\rightarrow\infty$ the slow mode dispersion tends to the Alfv{\'e}n mode dispersion which implies equipartition between kinetic and magnetic fluctuations. Also as $\beta$ increases the fraction of fast mode increases in the kinetic fluctuations, while decreasing in the magnetic fluctuations. This is expected from the fast mode eigenvector (Eq.~A30 of Ref.~\cite{ChoLazarian2003}) as in the high $\beta\rightarrow\infty$ limit we have $\bm{v}_{\bm{k},F}\propto \bm{k}$ and $\omega_{\bm{k},F}\sim kc_s$. This gives $|\bm{b}_{\bm{k},F}|\sim |\bm{v}_{\bm{k},F}|B_0/c_s$ and hence $|\bm{b}_{\bm{k},F}|/B_0\ll|\bm{v}_{\bm{k},F}|/v_A$.

If we consider only the velocity fluctuations then in all the compressively driven simulations, the total fraction of slow and fast modes is larger than the fraction of Alfv{\'e}n mode. Since the Alfv{\'e}n mode is incompressible while the fast and slow magnetosonic modes are compressible, this means that compressive driving expectedly makes the compressible velocity components dominant. If we consider only  the magnetic fluctuations, then in simulations C1a, C2a, and C4a, the slow plus fast fraction is larger than the Alfv{\'e}n fraction. Therefore, even magnetic fluctuations are dominated by the compressible magnetosonic modes in compressively driven turbulence of plasma $\beta$ close to unity. 

The mode fractions only give us a crude estimate of the strengths of various modes. We take a look at the perpendicular wavenumber energy spectrum to get a sense of the wavenumber distribution of the mode energies. This is shown in Fig.~\ref{1Dkspec_array}. The perpendicular direction is taken w.r.t. the $z$ direction which is the mean field direction. The spectrum in the perpendicular $x-y$ plane is averaged over the angle $\theta$ between $\bm{k}_{\perp}$ vector and $x$ axis, $E(k_{\perp})=\int d\theta k_{\perp} E(k_{\perp},\theta)$. This spectrum is very similar to the 1D wavenumber spectrum averaged over all three directions, $E(k)$. The $k_{z}=0$ wavenumber is removed from the data cube when calculating this spectrum in order to be consistent with Fig.~\ref{energy_fracs_barplot}. The energy spectra of the Alfv{\'e}n and slow modes show very similar behavior across all the cases. The solenoidally driven low $M_A$ case S1a is a case where the Alfv{\'e}n cascade is weak as we will see later. The fast mode in this case is very weak energetically compared to the slow and Alfv{\'e}n modes, with a very steep spectrum. In the solenoidally driven trans-Alfv{\'e}nic case S4a, the Alfv{\'e}n and slow modes show a spectrum close to $k_{\perp}^{-3/2}$, which is indicative of strong turbulence. The magnetic field spectrum of fast modes is close to $k_{\perp}^{-3/2}$ while the velocity field spectrum is between to $k_{\perp}^{-5/3}$ and $k_{\perp}^{-2}$. This is similar to earlier results of fast mode spectra~\citep{ChoLazarian2003}. 

In the compressively driven cases of C1a and C4a we see that the fast mode energy level increases compared to the solenoidally driven cases at large scales close to the driving scales. In C4a the Alfv{\'e}n and slow modes still show very similar spectra, close to $k_{\perp}^{-3/2}$. The fast mode velocity field in both C1a and C4a shows a spectrum of $k_{\perp}^{-2}$, while the magnetic field spectrum is between $k_{\perp}^{-5/3}$ and $k_{\perp}^{-2}$. This spectrum is steeper than the $k^{-3/2}$ spectrum claimed in Ref.~\cite{ChoLazarian2003}. Sharp jumps in data can lead to a steeper power spectrum~\cite{RobertsGoldstein1987} but might also limit the applicability of the mode decomposition.
We try to identify regions of these sharp gradients in a fast mode data cube of simulation C4a by finding cells which have velocity jumps in neighboring cells above a threshold. Regions with more than $30$\% jump in the velocity occupy 10\% of the total volume (more than 100\% jump regions occupy only 0.8\% volume). These sharp gradients are weak and occupy a very small volume therefore we expect the mode decomposition would still be valid for such data. To test this further we perform a test simulation with a superposition of 3 fast modes with mutually orthogonal wave-vectors which do not interact via radial 3-wave interactions. These modes steepen into shocks but the mode decomposition of the data still reveals the dominance of fast modes, as is expected. Therefore we can rely on mode decomposition even in this  scenario. As the slope of the fast mode spectrum is steeper, even though it is dominant close to the driving scale, its energy component drops off compared to the Alfv{\'e}n and slow modes at smaller scales.
%

\begin{figure}[t]
\centering  
\includegraphics[width=1.0\linewidth]{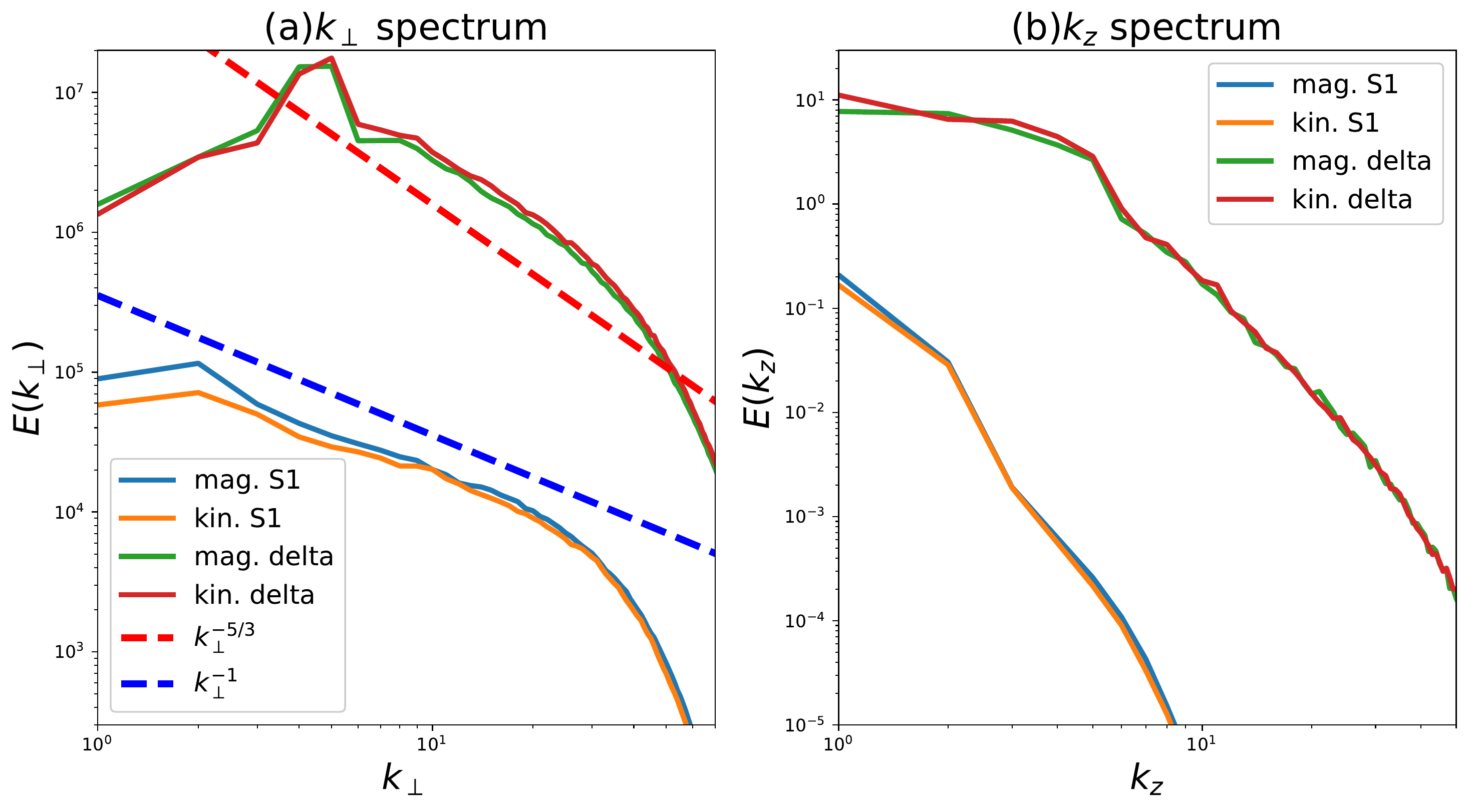}
\caption{(a) The $k_{\perp}$ spectrum excluding the $k_{z}=0$ mode. A delta-correlated in time forcing is used to produce a data cube with $M_A=0.24$ (labelled as `delta') and it is compared with the OU-process forced S1 simulation. The delta-correlated forcing produces a spectrum close to $k_{\perp}^{-5/3}$ while the S1 simulation has a spectrum $k_{\perp}^{-1}$. (b) Comparison of the $k_{z}$ spectrum. The OU forcing produces a very weak energy cascade to higher $k_{z}$ modes while the delta-correlated forcing produces a stronger cascade.}
\label{delta_forcing}
\end{figure}
The spectrum of Alfv{\'e}n and slow modes in simulations S1a and C1a shown in Fig.~\ref{1Dkspec_array} is shallower than $k_{\perp}^{-3/2}$ and closer to $k_{\perp}^{-1}$. These simulations were run for $T=16$ which is close to an order unity nonlinear time. We also ran the simulations for a longer time up to $T=32$ and verified that the spectra are converged with time. {As the forcing injects energy in the velocity field isotropically in a spherical shell of wavevectors with $1\le k\le 3$}, this could be a case of weak turbulence driven hydrodynamically at both $k_{z}\neq 0$ and $k_{z}=0$ {as considered in Ref.}~\cite{SchekochihinNazarenko2012}. {The Alfv{\'e}n modes appear with similar spectrum} in the S1 simulation. The spectrum of $k_{z}\neq 0$ modes is $k_{\perp}^{-1}$ as predicted (see Fig.~\ref{delta_forcing}a). The OU forcing plays an important role in producing this behavior. We implemented and tried the often-used delta-correlated in time forcing used in turbulence simulations like in Refs.~\cite{Brandenburg2001,BrandenburgDobler2001,ChoLazarian2002,ChoLazarian2003} and produced a turbulent date cube with $M_A=0.24$. We compare the spectrum produced by this simulation with the OU forced simulation S1a in Fig.~\ref{delta_forcing}. Theoretically a large range of weak turbulence is expected from $k_{inj}$ to $k_{inj}/M_{A}^{2}$~\cite{YanLazarian2008}. In the OU forced simulation, the $k_{z}$ spectrum is very steep indicating weak turbulence while the $k_{\perp}$ spectrum is $k_{\perp}^{-1}$. On the other hand the delta-correlated forcing produces a $k_{\perp}^{-5/3}$ spectrum which is more representative of strong turbulence along with a significant energy cascade to higher $k_{z}$ modes. The energy spectra of OU forced simulations in Fig.~\ref{delta_forcing} are also lower than the delta-correlated spectra since for the OU simulations more energy is concentrated in the $k_{z}=0$ modes. We think that the delta-correlated in time forcing produces faster dynamics whereas the OU forcing gives a slower evolution of the force allowing a weak cascade to develop.

\begin{figure}[h]
\centering 
\includegraphics[width=1.0\linewidth]{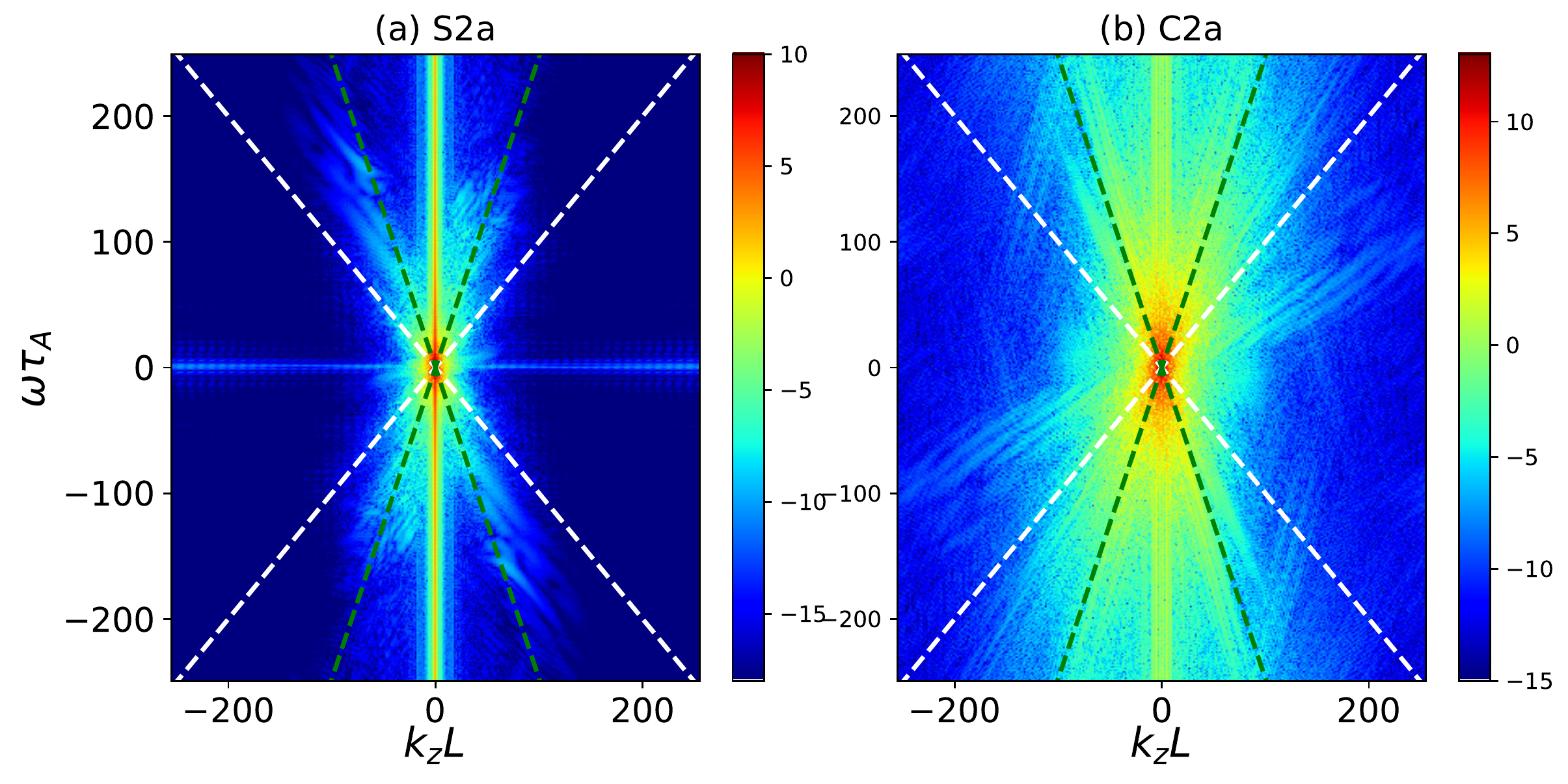}
\caption{Frequency spectra showing power density in $\omega$ versus $k_{z}$. The color represents the logarithm of power. The left plot is from velocity field of simulation S2a, while the right plot is for simulation C2a again using the velocity field. The white dashed line is a reference Alfv{\'e}n mode dispersion while the dashed green line is an example fast mode dispersion.}
\label{freq_spec_S2_C2}
\end{figure}
We can also see signatures of different modes in the frequency spectra. For this 1 dimensional slices along the $x$ (perpendicular to mean field) and $z$ (parallel to mean field) axes are taken. These slices are output at a high frequency at a time interval of $\Delta t=0.002\tau_A$. Then doing a 2D Fourier transform along the $x$($z$) and time axes gives us the power distribution in $k_{\perp}$($k_{z}$)-$\omega$ space. Fig.~\ref{freq_spec_S2_C2} shows this power spectrum in the $k_{z}-\omega$ space. The white dashed line follows the $\omega=\pm k_{z}v_A$ line, which is the dispersion relation of Alfv{\'e}n waves. The green dashed line traces the relation $\omega=(1+\sqrt{2})\sqrt{2}k_{z}v_A$. This is the relation for a fast mode where $c_s=v_A$ (approximately true for these simulations) and $k_{\perp}=k_{z}$. Fig.~\ref{freq_spec_S2_C2}(a) shows the frequency spectrum from the velocity fluctuations in simulation S2a, in which the Alfv{\'e}n and slow mode contributions dominate significantly over fast modes as seen in Fig.~\ref{energy_fracs_barplot}. We see that the power is concentrated close to the Alfv{\'e}nic dispersion. Fig.~\ref{freq_spec_S2_C2}(b) shows the frequency spectrum for velocity field in simulation C2a which also shows a branch of power concentrated at higher frequencies $\omega$ which are close to the fast mode dispersion. The fast mode is also a significant proportion of the mode mixture in C2a simulation (Fig.~\ref{energy_fracs_barplot}) and this reflects in the frequency characteristics. In the next section we focus on the anisotropy characteristics of the Alfv{\'e}n and slow modes.


\section{Alfv{\'e}n and slow modes}

\begin{figure}[h]
\centering 
\includegraphics[width=1.0\linewidth]{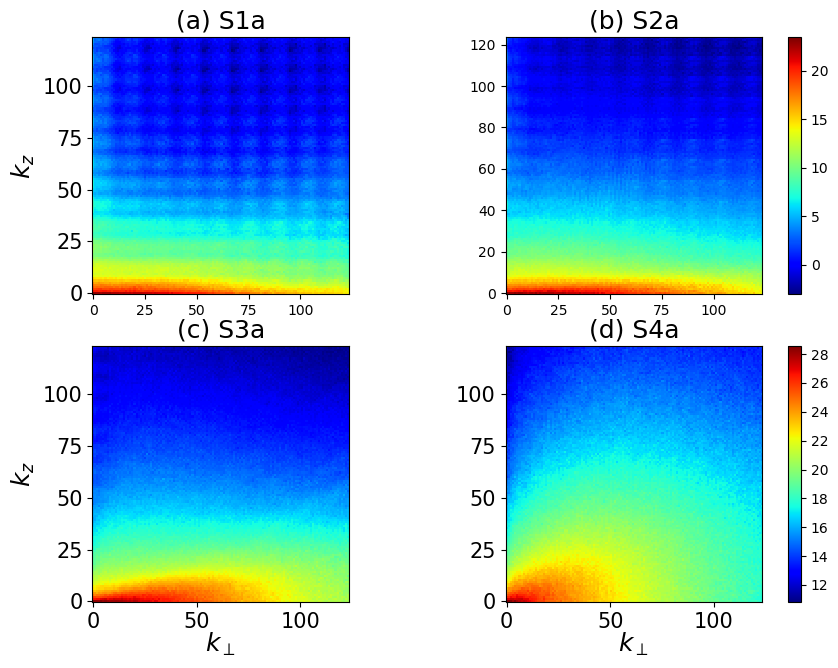}
\caption{The $k_{z}-k_{\perp}$ wavenumber spectrum for the velocity field of Alfv{\'e}n modes with increasing Mach number. The color indicates logarithm of the spectrum power. S1a is the lowest Mach number of 0.24, going up to S4a which has highest Mach number of 0.99. The power spreads more in the parallel direction as Alfv{\'e}n Mach number increases.}
\label{kspec_par_perp_A_S1-4}
\end{figure}

\begin{figure*}[t]
\centering 
\includegraphics[width=1.0\linewidth]{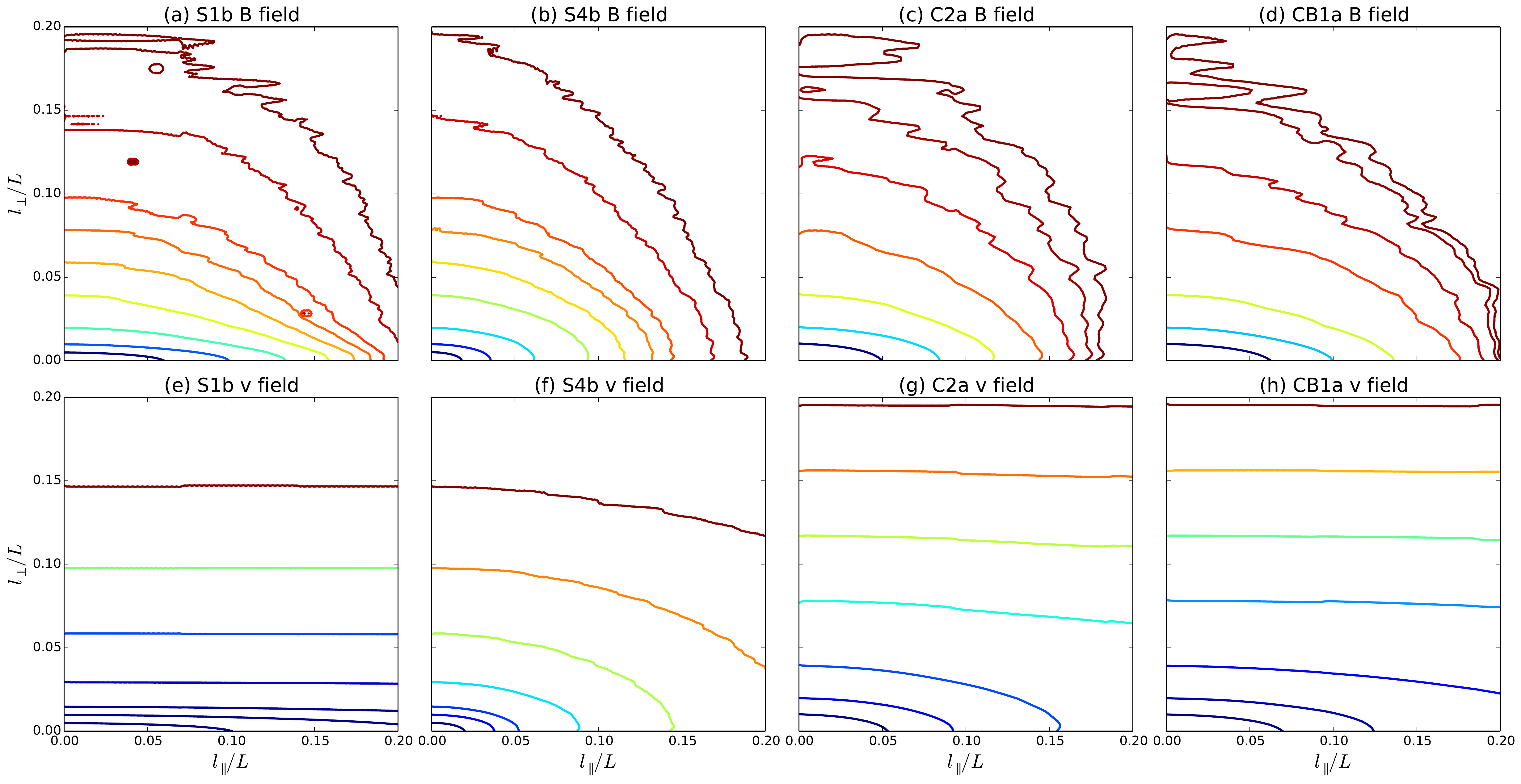}
\caption{Iso-contours of structure function for Alfv{\'e}n modes from their magnetic (top) and velocity fields (bottom) for the 4 simulations S1b, S4b, C2a, and CB1a. The units of $l_{\parallel}$ and $l_{\perp}$ are in terms of the simulation box size. We see the scale-dependent anisotropy for the Alfv{\'e}n modes which are longer in the parallel direction compared to the perpendicular direction.}
\label{sf2D_A}
\end{figure*}

We are interested in the nature of cascade of the different MHD modes, especially in its anisotropy. For this we analyze the energy spectrum in the $k_{z}$-$k_{\perp}$ space. Here the parallel direction is along the mean field, i.e. along the $z$ direction. The 2D spectrum is defined as $E(k_{z},k_{\perp})=\int k_{\perp}E(k_{\perp},\theta,k_z)d\theta$, where $k_z$ is parallel to the mean magnetic field, $k_{\perp}=\sqrt{k_x^2+k_y^2}$, and $\theta$ is the angle between $\bm{k}_{\perp}$ and $x$ axis. Fig.~\ref{kspec_par_perp_A_S1-4} shows these spectra for velocity field of the Alfv{\'e}n modes in simulations S1a to S4a with increasing $M_A$. We see that for S1a the energy is distributed along the $k_{\perp}$ axis close to $k_{z}=0$. There is very little cascade along the parallel direction in the sense that as energy spreads to higher $k_{\perp}$, there is no spread to higher $k_{z}$. The situation is similar for S2a simulation. As the $M_A$ increases the cascade slowly spreads in the parallel direction, slightly in the S3a simulation and more prominently in the S4a simulation. This is an indication of the turbulence transitioning from weak to strong as $M_A$ increases. 

We further analyze the anisotropy of the Alfv{\'e}n modes by using structure functions. The anisotropic structure function is defined as
\begin{align}
\label{sf_def}
SF_2(l_{\parallel},l_{\perp})=\langle |&\bm{b}(\bm{r}-(l_{\parallel}/2)\hat{\bm{b}}-(l_{\perp}/2)\hat{\bm{b}}_{\perp})- \nonumber \\
&\bm{b}(\bm{r}+(l_{\parallel}/2)\hat{\bm{b}}+(l_{\perp}/2)\hat{\bm{b}}_{\perp})|^2\rangle_{\bm{r}}.
\end{align}
This involves an ensemble average over a number of pairs of points which are separated by distance $l_{\parallel}$ in the magnetic field parallel direction ($\hat{\bm{b}}$) and distance $l_{\perp}$ in a field perpendicular direction ($\hat{\bm{b}}_{\perp}$). The magnetic field direction $\hat{\bm{b}}$ is the local mean magnetic field. To obtain this, first for each parallel and perpendicular distance pair ($l_{\parallel},l_{\perp}$) a distance $l=\sqrt{l_{\parallel}^2+l_{\perp}^2}$ is calculated. Then a random point is selected in the data cube and a sphere is taken around this point with a diameter $l$. The local mean magnetic field direction is calculated by taking an average of a few ($\geq 5$) random points located in this sphere. We have verified that the results do not change when taking more points. This gives us the local mean field direction $\hat{\bm{b}}$ and  a random unit vector perpendicular to $\hat{\bm{b}}$ is also constructed, $\hat{\bm{b}}_{\perp}$. This allows us to select 2 points on this sphere that are separated by $l_{\parallel}\hat{\bm{b}}+l_{\perp}\hat{\bm{b}}_{\perp}$. An ensemble average over thousands of such pairs gives us $SF_2(l_{\parallel},l_{\perp})$. 

The isocontours of the structure function are plotted in Fig.~\ref{sf2D_A}. A second order smoothing is applied a few times on the structure function to make the contours smoother, without changing their behavior. These are derived from the magnetic and velocity fields of the decomposed Alfv{\'e}n mode in the 4 different simulations S1b, S4b, C2a, and CB1a, averaged over several time snapshots. The driving occurs up to length scale of 0.33 units in terms of box length, and so we only focus on $l_{\parallel}$ and $l_{\perp}$ up to 0.2. The anisotropy of the Alfv{\'e}n modes is clearly visible across the different simulations and is quite similar. The parallel length scales are larger than the perpendicular length scales. Moreover, this ratio $l_{\parallel}/l_{\perp}$ changes with $l_{\parallel}$, with this anisotropy increasing as we proceed to smaller scales making this a scale-dependent anisotropy. Comparing S4b and C2a simulations shows that the anisotropy of Alfv{\'e}n modes is not affected by the type of driving. Changing the plasma $\beta$ in simulation CB1a also does not seem to change the anisotropy. The simulation S1b shows a structure function of the velocity field that is highly elongated along the $l_{\parallel}$ direction. Similar behavior is seen for the larger $l_{\perp}$ values in C2a and CB1a velocity field structure functions. These are low $M_A$ simulations. We observe that at low $M_A$ the velocity field is highly correlated along the magnetic field direction, giving rise to an almost $l_{\parallel}$-invariant structure function. This is indicative of a weak nature of cascade at low $M_A$. As $M_A$ increases smaller-scale structure develops in the parallel direction. The trans-Alfv{\'e}nic simulation S4b shows structure function that has smaller $l_{\parallel}$ scales. We need to look further at the variation of $l_{\parallel}$ with $l_{\perp}$ to get a quantitative understanding of the anisotropy of Alfv{\'e}n modes.

We also analyzed the 2D structure functions of the slow modes. They show a behavior similar to the Alfv{\'e}n modes with an anisotropy showing $l_{\parallel}>l_{\perp}$ at all scales.  

\begin{figure}[t]
\centering 
\includegraphics[width=1.0\linewidth]{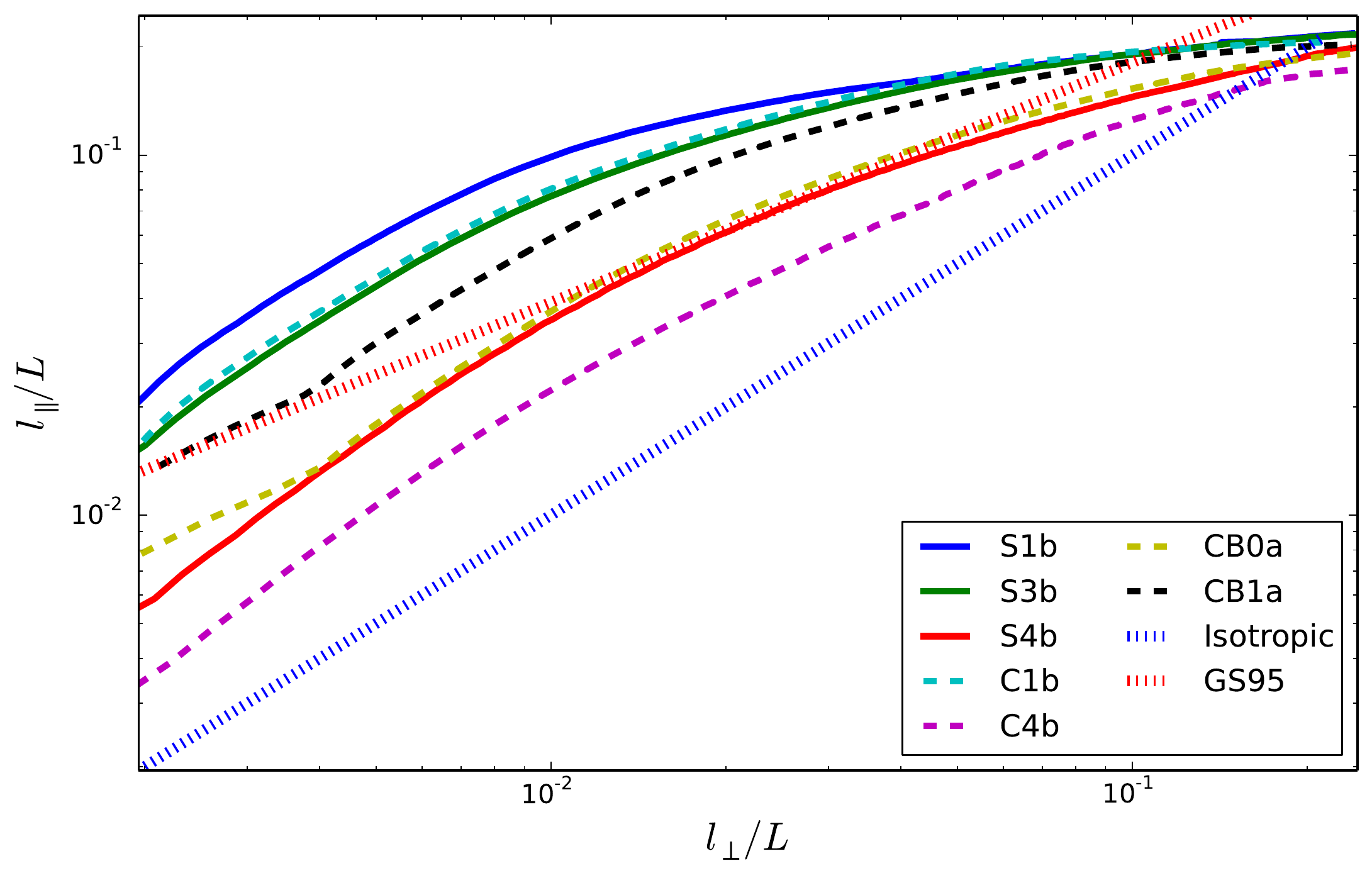}
\caption{The variation of $l_{\parallel}$ versus $l_{\perp}$ for the Alfv{\'e}n modes magnetic field for the different simulations. The Goldreich-Sridhar scaling is $l_{\parallel}\sim l_{\perp}^{2/3}$ shown by the red dotted line (GS). Isotropic behavior $l_{\parallel}=l_{\perp}$ is shown by the blue dotted line.}
\label{lpar_lperp_A}
\end{figure}

\begin{figure}[t]
\centering 
\includegraphics[width=1.0\linewidth]{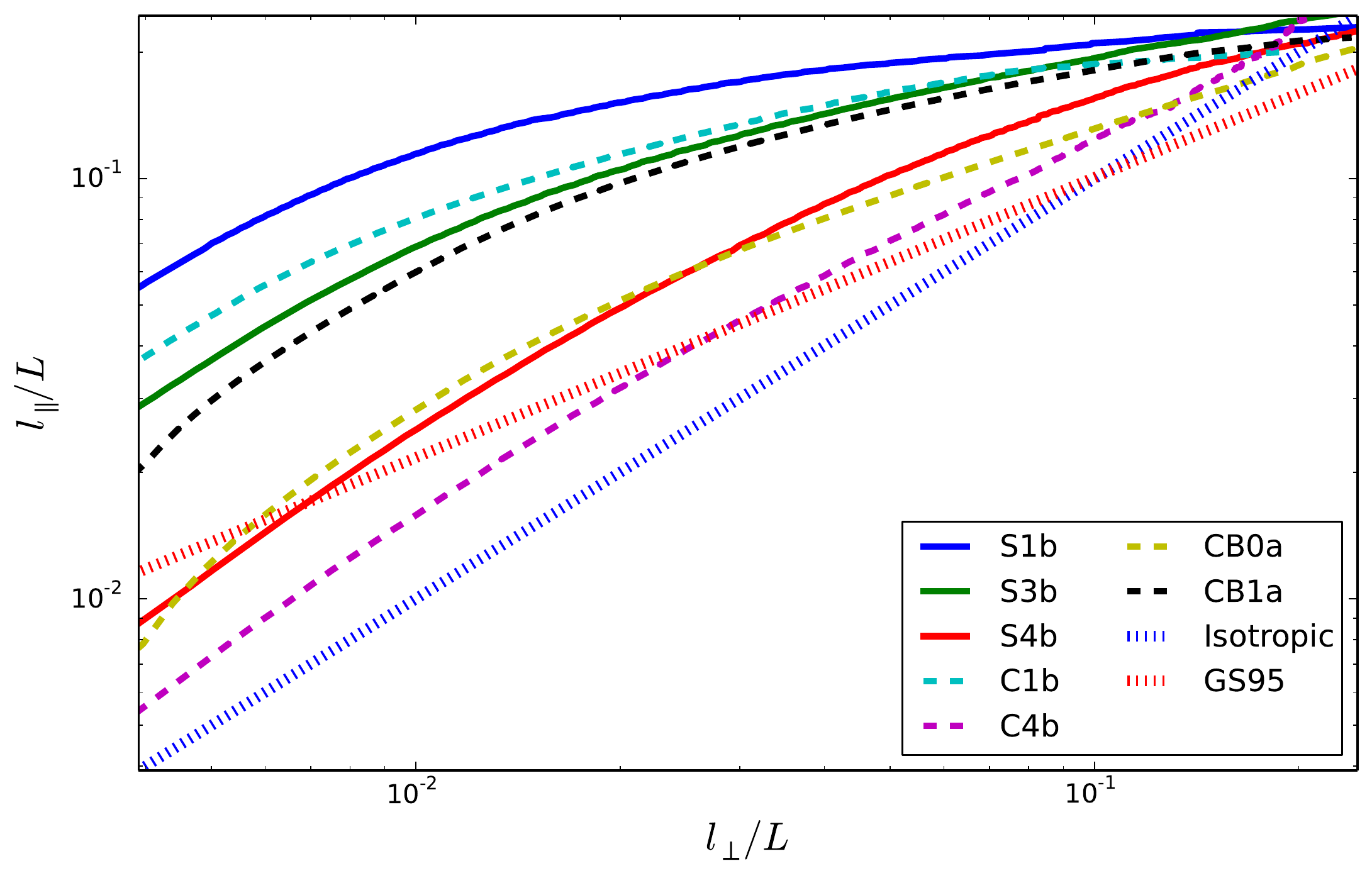}
\caption{The variation of $l_{\parallel}$ versus $l_{\perp}$ for the slow modes magnetic field for the different simulations. The Goldreich-Sridhar scaling is $l_{\parallel}\sim l_{\perp}^{2/3}$ shown by the red dotted line (GS). Isotropic behavior $l_{\parallel}=l_{\perp}$ is shown by the blue dotted line.}
\label{lpar_lperp_S}
\end{figure}

From the 2D structure functions we can extract the relation of $l_{\parallel}$ versus $l_{\perp}$ by taking the cuts of the isocontours of constant structure function at the $l_{\parallel}$ and $l_{\perp}$ axis. For this the structure function along the two axis ($l_{\parallel}=0$ and $l_{\perp}=0$) is calculated with very high statistics such that a smooth interpolation can be used to obtain a relation between the $l_{\parallel}$ and $l_{\perp}$. This measure gives a better picture of the anisotropy scaling, shown in Fig.~\ref{lpar_lperp_A}. It shows this measure derived from the magnetic fields of the Alfv{\'e}n modes. We see that the anisotropy depends on the $M_A$ for the Alfv{\'e}n modes. For simulation S1b the $l_{\parallel}$ is almost constant (very weakly changing) at large $l_{\perp}$ implying that eddies form at smaller perpendicular length scales but maintain the same parallel length scales, which is an indication of weak turbulence. For the simulation S3b also a similar behavior is observed at large $l_{\perp}$, but the $l_{\parallel}$ starts reducing as $l_{\perp}$ gets smaller. As the Mach number increases in simulation S4b, the $l_{\parallel}$ scales close to the Goldreich-Sridhar scaling of $l_{\parallel}\sim l_{\perp}^{2/3}$. The C4b simulation shows a power law with the Goldreich-Sridhar scaling for a significant range of $0.02\lesssim l_{\perp}\lesssim 0.1$. Comparing S1b and C1b shows not much difference between solenoidal and compressive driving. Comparing CB0a with CB1a shows that the plasma $\beta$ does not have a significant effect on the anisotropy. This shows that at low $M_A$ there is a large range of scales in $l_{\perp}$ where $l_{\parallel}$ remains unchanging with $l_{\perp}$, indicating no cascade in parallel direction. As $M_A$ increases, this range decreases and we start seeing a transition to the Goldreich-Sridhar scaling.

Fig.~\ref{lpar_lperp_S} shows the variation of $l_{\parallel}$ with $l_{\perp}$ derived from the magnetic fields of the slow modes. It shows a behavior similar to the Alfv{\'e}n modes. For simulations S1b, S3b, C1b, and CB1a, the $l_{\parallel}$ drops very slowly with decreasing $l_\perp$, which is similar to the weak nature of Alfv{\'e}n modes in Fig.~\ref{lpar_lperp_A}. Simulations CB0a, C4b, and S4b show a behavior close to Goldreich-Sridhar scaling. There is some isotropic scaling also seen in simulation C4b. This could be due to coupling with the fast mode which is strong in this case and shows isotropic scaling as we will see later.
 
\begin{figure}[h]
\centering 
\includegraphics[width=1.0\linewidth]{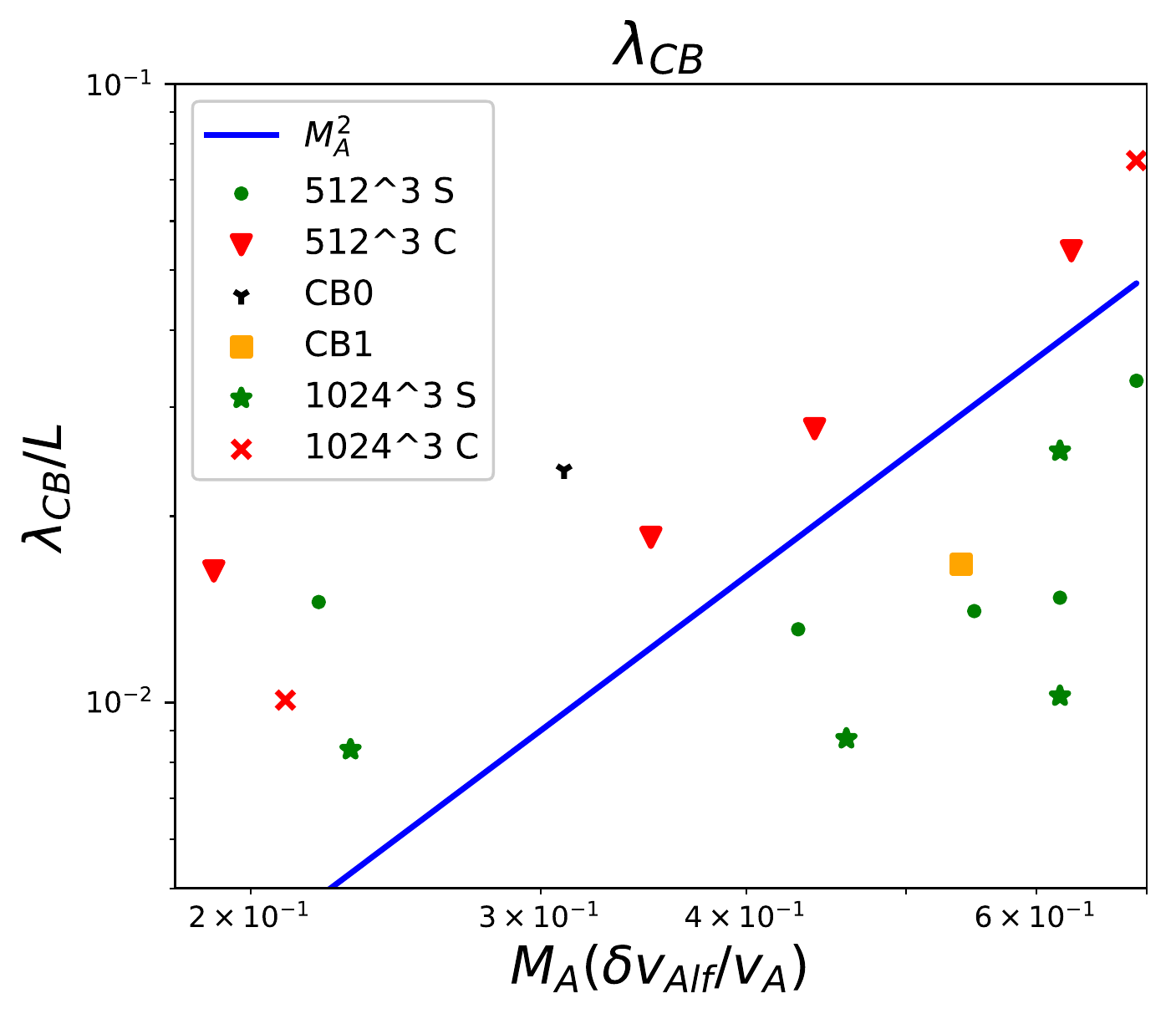}
\caption{The variation of the estimated transition scale of weak to strong turbulence $\lambda_{CB}$ with the Alfv{\'e}nic Mach number $M_A$. The different markers are the results from the different simulations where S stands for solenoidal and C for compressive runs. The Alfv{\'e}nic Mach number is calculated from the decomposed Alfv{\'e}n data. The blue line is showing the $M_A^2$ reference line.}
\label{lambdaCB_ma}
\end{figure}

The Alfv{\'e}n cascade is expected to transition from weak to strong at a transition scale $\lambda_{CB}$. The weak turbulence spectrum of Alfv{\'e}n modes is $E(k_{\perp})\sim (\epsilon/\tau_{A})^{1/2}k_{\perp}^{-2}$~\citep{BoldyrevPerez2009,SchekochihinNazarenko2012}, where $\epsilon$ is the energy injection rate. Then, $\lambda_{CB}$ is the scale where the linear interaction time, $\tau_{A}=L/v_A$, balances the nonlinear interaction time, $\tau_{nl}=\lambda_{CB}/\delta v_{\lambda_{CB}}$. From the weak turbulence spectrum, the velocity strength goes as $\delta v_{\l_{\perp}}\sim l_{\perp}^{1/2}$. If we assume a velocity field of strength $v_{A}M_A$ at injection scale $L$, then $\delta v_{\lambda_{CB}}\sim M_Av_A({\lambda_{CB}}/L)^{1/2}$. Balancing the linear and nonlinear interaction times gives $\lambda_{CB}\sim LM_{A}^{2}$~\cite{Kulsrudbook,YanLazarian2008}. Thus, when $M_A\lesssim1$, the weak regime of Alfv{\'e}nic turbulence should exist in the range of scales $[LM_{A}^2,L]$ while at smaller scales it should be in the strong regime. 

We try to estimate this transition scale by making use of the structure function anisotropy. As seen in Fig.~\ref{lpar_lperp_A} the $l_{\parallel}$ is expected to be invariant as a function of $l_{\perp}$ in the weak turbulence regime (at large $l_{\perp}$) and tend towards the Goldreich-Sridhar slope of $l_{\parallel}\sim l_{\perp}^{2/3}$ in the strong regime. Therefore, we fit a power low of the form $l_{\parallel}=Cl_{\perp}^{\alpha}$ at each $\l_{\perp}$ in a window of $l_{\perp}-\Delta$ to $l_{\perp}+\Delta$ around it. We take $\Delta=5$ grid points and the scaling exponent $\alpha$ is calculated at each $l_{\perp}$. As expected at large scales close to the driving scale the exponent is very close to 0 and it increases as $l_{\perp}$ reduces. It crosses the Goldreich-Sridhar value of $2/3$ for the first time at some $l_{\perp}$ value which can be taken as the transition scale $\lambda_{CB}$. So starting from large $l_{\perp}$ as we go down to smaller $l_{\perp}$, we define the transition scale $\lambda_{CB}$ as the $l_{\perp}$ at which $\alpha$ goes over a threshold value of $2/3$ for the first time. This way $\lambda_{CB}$ is identified for the Alfv{\'e}n mode in all the different simulations. The variation of this transition scale as a function of $M_A$ for each of these simulations is shown in Fig.~\ref{lambdaCB_ma}. As the weak to strong transition is expected for the Alfv{\'e}n cascade, we take the $M_A$ of the decomposed Alfv{\'e}n mode instead of that of the total data. It is broadly indicative of the $M_A^2$ dependence as expected from theory. At low $M_A$ the $\lambda_{CB}$ is already on scales close to the dissipative scales. Therefore, we see a plateau at low $M_A$. However, the trend is clearer from the simulations with {$M_A\gtrsim 0.4$}. {Although there are only a few points with significant scatter and there also appears to be some systematic variation of the transition scale with plasma $\beta$ and type of forcing, power law fits to these points are close to $M_{A}^{2}$ scaling.} This is an important verification of the existence of the weak regime of Alfv{\'e}n turbulence. This feature needs to be taken into account when developing models of interstellar medium turbulence for cosmic ray scattering. Next we focus on the properties of the fast mode.

\section{Fast modes}

\begin{figure}[h]
\centering 
\includegraphics[width=1.0\linewidth]{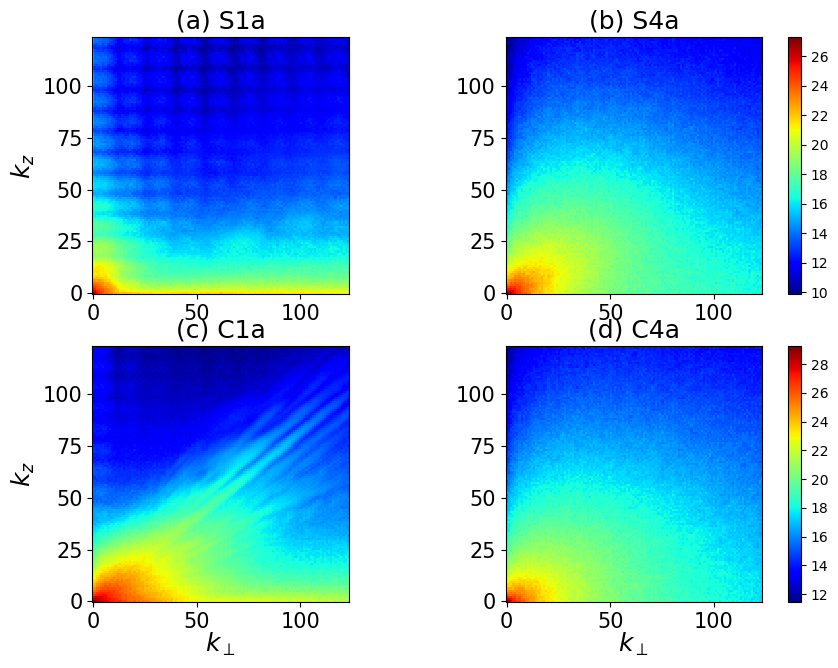}
\caption{$k_{z}-k_{\perp}$ spectrum of the velocity field of fast modes for the 4 simulations S1a, S4a, C1a, and C4a.}
\label{2Dkspec_F}
\end{figure}

In Fig.~\ref{1Dkspec_array} the fast modes showed a different spectrum compared to the Alfv{\'e}n and slow modes. Fig.~\ref{2Dkspec_F} shows the fast mode wavenumber spectrum as a function of $k_{\perp}$ and $k_{z}$. Contrasting it with Fig.~\ref{kspec_par_perp_A_S1-4} we see a different nature of the cascade here. For the Alfv{\'e}n mode the cascade was clearly anisotropic with energy cascade taking place mostly in the direction of larger $k_{\perp}$. However, the spread of energy for the fast mode appears very close to isotropic as there is almost uniform distribution of power in the parallel and perpendicular wavenumbers. In the low-$M_A$ case of S1a where the fast mode has a tiny fraction, the energy distribution seems isotropic at low $k$ with a small anisotropic cascade along the $k_{\perp}$ direction for higher $k_{\perp}$. However, for the other three simulations S4a, C1a, and C4a, the cascade is extending radially. Another aspect is that the cascade for Alfv{\'e}n mode is changing with $M_A$ in Fig.~\ref{kspec_par_perp_A_S1-4}, with almost no cascade in the parallel direction for S1a case due to the weak nature of turbulence. Here in both C1a and C4a simulations the fast mode shows an isotropic cascade. The isotropic nature of the fast mode cascade is similar even with solenoidal and compressive driving. This shows that the isotropic nature of fast mode cascade is a robust feature. 

\begin{figure*}[t]
\centering 
\leavevmode 
\includegraphics[width=1.0\linewidth]{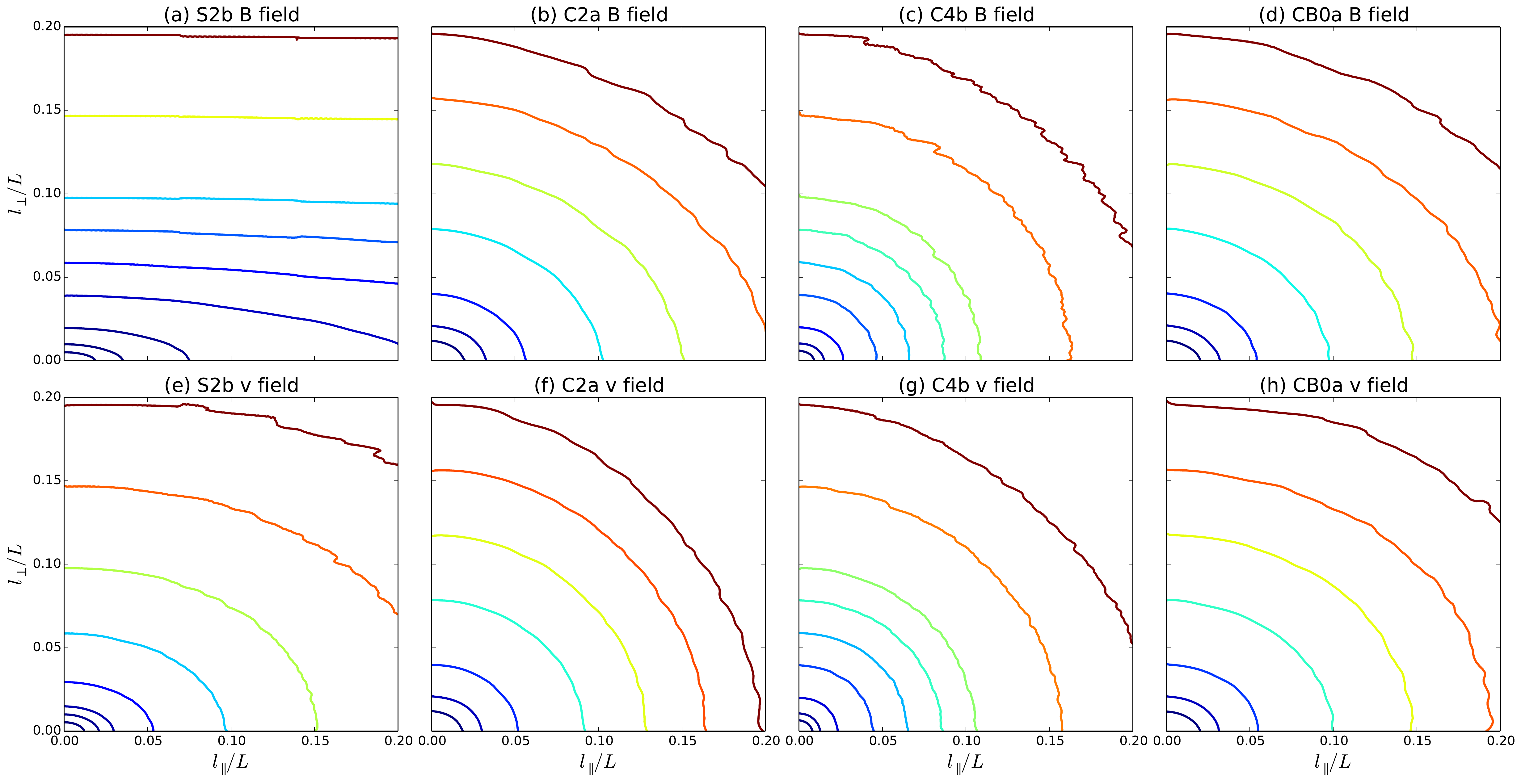}
\caption{2D structure functions for the fast mode for magnetic fields (top) and velocity fields (bottom) for the simulations S2b(a,e), C2a(b,f), C4b(c,g), and CB0a(d,h). It seems isotropic for the cases C2a, C4b, and CB0a while for S2b there seems to be some anisotropy.}
\label{sf2D_F}
\end{figure*}

\begin{figure}[t]
\centering 
\includegraphics[width=1.0\linewidth]{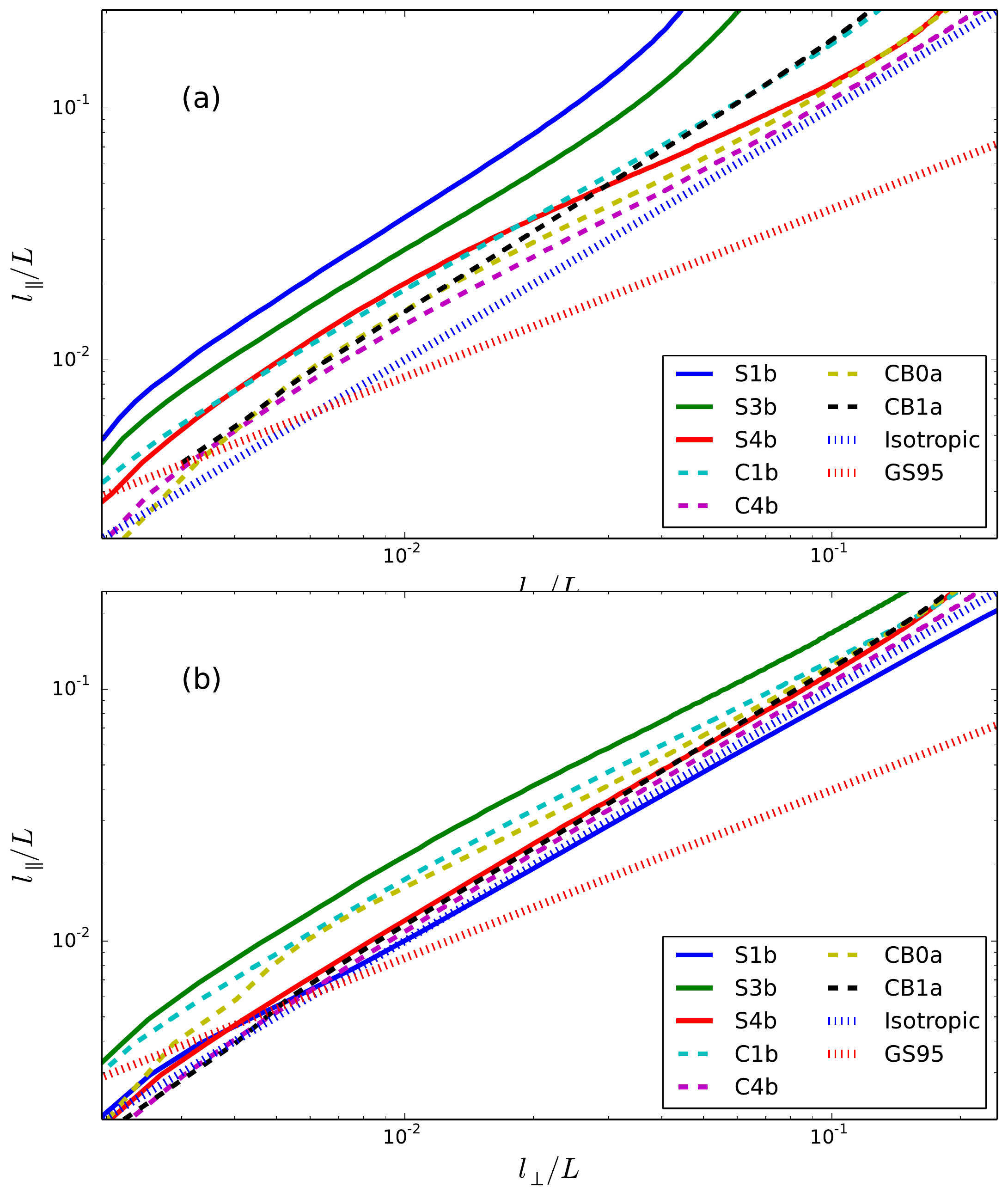}
\caption{The variation of $l_{\parallel}$ versus $l_{\perp}$ for the fast modes for the (a) magnetic field and (b) velocity field. The blue dotted line shows isotropic behavior, while the red dotted line shows the Goldreich-Sridhar scaling.}
\label{lpar_lperp_B_v_F}
\end{figure}

\begin{figure}[t]
\centering 
\includegraphics[width=1.0\linewidth]{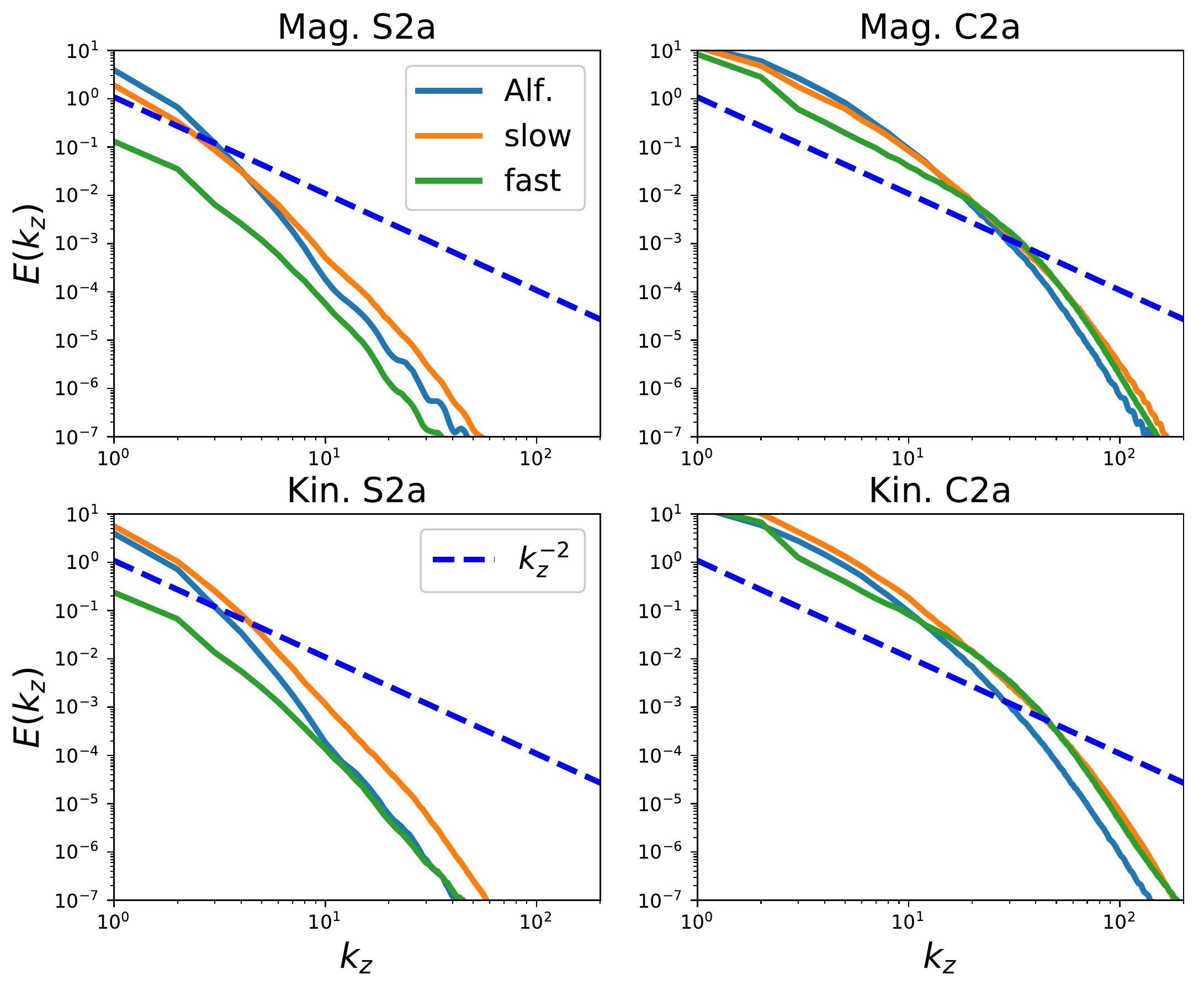}
\caption{The parallel wavenumber spectrum for the three modes for S2a simulation (left) and C2a simulation (right), with magnetic fields (top) and velocity fields (bottom). The legends apply to all panels.}
\label{kspeckpar_array}
\end{figure}

The 2D structure function isocontours of the velocity and magnetic fields of fast modes are shown in Fig.~\ref{sf2D_F}. These contours also show the isotropic nature for the various simulations. For the C2a and C4b simulations, we can see the isotropic contours extending almost up to $l_{\perp}/L\sim 0.2$ which is close to the driving scale of $0.3$. Also in the $1024^3$ simulations, this isotropy extends well up to scales smaller than $l_{\perp}/L\sim 0.05$. From these results we can say that the isotropy does indeed extend throughout the inertial range. The magnetic field contours in C2a show a slight anisotropy at large $l_{\parallel}$. However, the contours from the velocity field are highly isotropic. From Fig.~\ref{energy_fracs_barplot} we know that in simulation C2a, the fast mode fraction is larger in the velocity field than in the magnetic field, so the velocity field shows a clearer isotropy. Cases CB0a and S2b show a little anisotropy similar to Alfv{\'e}n modes. These are cases where the fast mode fraction is smaller (Fig.~\ref{energy_fracs_barplot}) compared to other cases. This indicates that the properties of a mode might be influenced by how strong it is compared to the other modes. If it is weak then its properties might get influenced by the dominant mode properties. This points to the cascades of these modes not being entirely independent of each other, with some interaction between them. 


We calculate the variation of $l_{\parallel}$ with $l_{\perp}$ for the fast modes now. Fig.~\ref{lpar_lperp_B_v_F}(a) shows this for the fast mode magnetic field while Fig.~\ref{lpar_lperp_B_v_F}(b) is from the velocity field. The magnetic field of the simulations S1b, S3b, C1b,CB0a, C4b, and CB1a follows a scaling very close to the isotropic scaling. In simulation S4b the relation is closer to GS95 scaling, but this is a case of weak fast modes.  For the velocity field the relationship $l_{\parallel}=l_{\perp}$ is followed very closely in almost all the simulations. This verifies that the isotropic nature of fast modes is a robust feature and it extends throughout the inertial range. It is the same for both low and high $M_A$, as well as high $\beta$ and low $\beta$, as long as the driving generates a significant fraction of fast modes.


Considering the fact that the cascade of energy for Alfv{\'e}n and slow modes is stronger in the perpendicular direction while for the fast mode it is isotropic, it is interesting to compare their spectrum in the parallel direction. Fig.~\ref{kspeckpar_array} compares their parallel spectra for the cases S2a and C2a. In the case of S2a the cascade of all three modes is very weak in the parallel direction as their spectra are very steep compared to the $k_{z}^{-2}$ reference slope. In the case of C2a, the Alfv{\'e}n and slow modes are still expected to have a weak cascade. However, the fast mode will be more energetic since this is compressively driven and will be isotropic. Thus, the fast mode parallel spectrum shows a $k_{z}^{-2}$ slope while the Alfv{\'e}n and slow modes on the other hand retain a weak cascade with a spectrum steeper than $k_{z}^{-2}$. As a result, even at high $k_{z}$ ($k_{z}>20$ in Fig.~\ref{kspeckpar_array}) the fast mode has a slightly greater energy that the Alfv{\'e}n and slow modes. Thus, in this regime of compressively driven, low $M_A$, weak Alfv{\'e}n turbulence, the fast mode can be dominant to the Alfv{\'e}n and slow modes even at large parallel wavenumbers, which has implications for the cosmic ray scattering and transport by the turbulence since the gyroresonance condition is set by the parallel wavenumber \cite{YanLazarian2002}. 


\section{Astrophysical Implications}

This work studies the role of forcing in shaping MHD turbulence, in particular, by analyzing the decomposed linear MHD eigenmodes and their properties. The decomposition was introduced earlier to study the behavior of each eigenmode in incompressively driven turbulence \cite{ChoLazarian2002}. We are interested in the regime of $M_A \lesssim 1$ where the decomposition in the Fourier space provides an effective instrument to reveal some intrinsic properties of MHD turbulence. By default, turbulence enters MHD regime only when $M_A\lesssim 1$. Super-Alfv{\'e}nic turbulence is hydrodynamic down to the scale $L/M_A^3$, where $\delta v/v_A$ reaches unity. Validity of mode decomposition is further confirmed here by the distinctive 3D characteristics of each eigenmode observed in different turbulence data cubes, resulting from both solenoidally and compressively driven MHD simulations. Several observational studies also show that sub or trans-sonic turbulence is common in astrophysical plasmas. Turbulence in the HII ionized gas of Orion nebula has been observed to be sub-sonic~\cite{ArthurMedina2016}. Measurements of cold and dense gas in an infrared dark cloud (IRDC) have also shown many regions of sub-sonic Mach number turbulence~\cite{SokolovWang2018}. Observations of the hot gaseous atmosphere of a nearby elliptical galaxy has also shown sub-sonic Mach number turbulence~\cite{OgorzalekZhuravleva2017}.

The simulations in this work use a form of forcing which can be easily decomposed into solenoidal and compressive parts. Any continuous, smooth forcing field confined in a limited region of space can always be Helmholtz decomposed into solenoidal and irrotational fields. In this sense, the forcing considered here is general. Supernova shocks and jets would drive turbulence compressively. On the other hand magneto-rotational instability and shear-driven turbulence is probably solenoidal. There could also be mixed regions of turbulence in astrophysical jets, with the shear flows that develop between their inner and outer parts driving solenoidal modes while the bow shock at the tip of the jet driving compressive modes~\cite{Federrath2018}. Thus one can imagine different neighboring regions of space with different types of driving and turbulence. 

We found that the OU forcing provides a more realistic forcing on large scales which can also reproduce weak turbulence features. An important demonstration of this work is the transition from weak to strong Alfv{\'e}nic turbulence. This implies that even if the turbulence is weak at injection scales, at scales smaller than $L_{inj}M_A^2$, turbulence will transition to a strong regime of Alfv{\'e}nic turbulence which shows the Goldreich-Sridhar anisotropy. This is important for the turbulent reconnection scenario proposed in Ref.~\cite{LazarianVishniac1999}. {Turbulence is closely connected with magnetic reconnection. The plasmoid instability can lead to a faster rate of reconnection~\cite{BhattacharjeeHuang2009,HuangComisso2017}. Mean field modeling shows that a balance between turbulence-driven transport enhancement and suppression can also lead to fast reconnection~\cite{YokoiHigashimori2013,WidmerBuchner2019}}. In the turbulence reconnection scenario of Ref.~\cite{LazarianVishniac1999} the transverse Alfv{\'e}n mode perturbations cause field line wandering which modifies the Sweet-Parker reconnection rate by a factor of $(L/l_{\parallel})^{1/2}$, where $L$ is the total length of the reconnection layer along the magnetic field while $l_{\parallel}$ is the parallel length of a turbulent eddy. Strong Alfv{\'e}nic turbulence will generate smaller lengths $l_{\parallel}$ according to the GS95 scaling and this will reduce the size of the local reconnection zone, increasing the speed of reconnection. On the other hand, weak Alfv{\'e}nic turbulence will not generate smaller $l_{\parallel}$ and won't speed up the reconnection rate. If the Alfv{\'e}n mode is dominant then below the transition scale we will always have strong turbulence and hence fast turbulent reconnection as a result. The magnetic field fluctuations in compressible modes are mostly in the parallel direction. Theoretically this will reduce the field line wandering and might not produce as much fast reconnection as in Alfv{\'e}n turbulence if compressible modes dominate in turbulence. In reality, the decrease is limited since the energy fraction in Alfv\'en modes does not vary beyond a factor of two (see Fig. 2).

Turbulence leads to transport and diffusion of cosmic rays. The different turbulence modes have different scattering and transport properties of cosmic rays~\cite{YanLazarian2002,YanLazarian2004, LynnQuataert2014}. The compressible modes are demonstrated to dominate the transport and acceleration in turbulence. The scattering by Alfv\'enic turbulence is substantially reduced due to the scale dependent anisotropy. A natural consequence of having different types of turbulence in different regions of space would be an inhomogeneous transport and diffusion of cosmic rays. In this work we have identified a regime of low-$M_A$ compressively driven turbulence where the isotropic fast mode can dominate over the Alfv{\'e}n mode in the  $k_{z}$ spectrum, which is the relevant one for gyroresonance of particles ($\omega-k_{z}v_{\parallel}=n\Omega$, with $\Omega$ the gyro-frequency). Though appearing to be weak, this turbulence still could scatter cosmic rays efficiently due to the large proportion of fast modes. Studying this regime of turbulence and its implications for cosmic ray distribution is thus important.

In the regime of strong Alfv{\'e}nic turbulence diffusion of particles in turbulent magnetic fields follows super-diffusion at scales smaller than the injection scales, $l_{\perp}^2\sim s^3$, where $l_{\perp}$ is the average separation between particles in the perpendicular direction, while $s$ is the distance traversed along the magnetic field lines~\cite{Richardson1926,LazarianYan2014}. This is because particles travel along magnetic field lines and the field separation grows as the $3/2$ power of the distance traveled along field lines. This super-diffusion behavior will last till the Alfv{\'e}n turbulence remains strong and the particles don't scatter. From this work it can be confirmed that in case of $M_A\lesssim 1$ turbulence, particle trajectories will separate super-diffusively at small scales until they reach the transition scale $L_{inj}M_A^2$. From then onwards they will be in a normal diffusion regime. A larger fraction of the compressible modes in the turbulence can change the field line separation rate and also the mean free path of particles, which will result in a slower superdiffusion.

\section{Summary}

We have analyzed the properties of the different MHD modes in solenoidally and compressively driven turbulence which is highly relevant for astrophysical plasmas. One of the important aims of studying different MHD modes is for their different properties in particle acceleration and  scattering, so these modes have been studied separately in this context. But an important open question was what are the relative fractions of these modes in astrophysical plasma turbulence and what do they depend on. 


Here we found that the type of driving is a major factor affecting the composition of these modes. Understandably, compressive driving leads to a larger proportion of the slow plus fast magnetosonic modes, but more specifically the fast magnetosonic modes. Moreover, the compressive modes can also dominate in the magnetic field fluctuations. This is crucial for the cosmic ray transport and acceleration in turbulence since fast modes dominate the particle scattering~\cite{YanLazarian2002,YanLazarian2004}. A signature from polarization analysis (SPA) method  to identify such modes from synchrotron polarization maps has been invented and this study holds important implications for such measurements~\cite{ZhangChepurnov2019}. 

The nature of turbulent cascade is also a highly relevant feature for a variety of astrophysical phenomena. We identify many important properties of the Alfv{\'e}n and slow mode turbulent cascade from these simulations. We see the anisotropic nature of the cascade as predicted by Goldreich-Sridhar theory. We also see indications of a weak regime of Alfv{\'e}nic turbulence that extends from $L_{\text{inj}}$ up to $L_{\text{inj}}M_A^2$. This is an important numerical test of a theoretically predicted regime.

On the other hand, the isotropic nature of the fast mode cascade which was earlier seen in low resolution studies is now verified to extend throughout the inertial range through higher resolution studies. This seems a robust feature which does not show a weak-strong transition unlike the Alfv{\'e}n and slow modes and does not depend on plasma $\beta$ or $M_A$. The spectrum of the fast modes can be steeper than $k^{-3/2}$ and closer to $k^{-2}$ when the fast mode dominates. This has implications for the cutoff scale and damping of fast modes.


We argue that the mode decomposition analysis will be relevant in regions of astrophysical plasmas smaller than the driving scales of turbulence with $M_A\lesssim 1$ and devoid of strong gradients. It implies the wide applicability of fast turbulent reconnection and superdiffusion of particle trajectories below the transition scale to strong Alfv{\'e}nic turbulence. Differently driven turbulence regions will lead to inhomogeneity in particle transport and acceleration. In the case of low $M_A$, compressively driven turbulence, the energy in fast modes becomes dominant even on small parallel scales. Thus cosmic ray scattering and acceleration will remain effective counterintuitively in this regime through both the gyroresonance and transit-time damping interactions with fast modes.


\begin{acknowledgments}
The authors acknowledge the North-German Supercomputing Alliance (HLRN) for providing HPC resources that have contributed to the research results reported in this paper.
\end{acknowledgments}

\bibliography{draft13}

\end{document}